\documentclass[preprint,5p,times,twocolumn]{elsarticle}

\usepackage{graphicx}
\usepackage{amssymb}

\usepackage{longtable}   
\usepackage{float} %allows the use of [H] to force positioning of figures
\usepackage{subfig}
\usepackage{hyperref} %used in references now  --- it shoud be removed
\usepackage{booktabs}
\usepackage{tabularx}

\journal{Information and Software Technology}

\begin{document}

\begin{frontmatter}

%% Title, authors and addresses

%% use the tnoteref command within \title for footnotes;
%% use the tnotetext command for the associated footnote;
%% use the fnref command within \author or \address for footnotes;
%% use the fntext command for the associated footnote;
%% use the corref command within \author for corresponding author footnotes;
%% use the cortext command for the associated footnote;
%% use the ead command for the email address,
%% and the form \ead[url] for the home page:
%%
%% \title{Title\tnoteref{label1}}
%% \tnotetext[label1]{}
%% \author{Name\corref{cor1}\fnref{label2}}
%% \ead{email address}
%% \ead[url]{home page}
%% \fntext[label2]{}
%% \cortext[cor1]{}
%% \address{Address\fnref{label3}}
%% \fntext[label3]{}

\title{Software Development in Startup Companies: A Systematic Mapping Study}

%% use optional labels to link authors explicitly to addresses:
%% \author[label1,label2]{<author name>}
%% \address[label1]{<address>}
%% \address[label2]{<address>}

%\tnotetext[bth]{Blekinge Institute of Technology}

\author[bth]{Nicol\`{o} Paternoster}
\author[bth]{Carmine Giardino}
\author[bth]{Michael Unterkalmsteiner}
\author[bth]{Tony Gorschek}
\author[ubz]{Pekka Abrahamsson}

\address[bth]{ Blekinge Institute of Technology, SE-371 79 Karlskrona, Sweden}
\address[ubz]{Free University of Bolzano-Bozen, I-39100 Bolzano-Bozen, Italy}

\begin{abstract}
\small
\textit{Context:} Software startups are newly created companies with no
operating history and fast in producing cutting-edge technologies.
These companies develop software under highly uncertain conditions, tackling
fast-growing markets under severe lack of resources. Therefore, software 
startups present an unique combination of characteristics which pose several 
challenges to software development activities. 
\textit{Objective:} This study aims to structure and analyze the literature on
software development in startup companies, determining thereby the potential 
for technology transfer and identifying software development work practices 
reported by practitioners and researchers.
\textit{Method:} We conducted a systematic mapping study, developing a 
classification schema, ranking the selected primary studies according their 
rigor and relevance, and analyzing reported software development work practices 
in startups.
\textit{Results:} A total of 43 primary studies were identified and mapped, 
synthesizing the available evidence on software development in startups. Only 
16 studies are entirely dedicated to software development in startups, of which 
10 result in a weak contribution (advice and implications (6); lesson learned 
(3); tool (1)). Nineteen studies focus on managerial and organizational 
factors. Moreover, only 9 studies exhibit high scientific rigor and relevance. 
From the reviewed primary studies, 213 software engineering work practices were 
extracted, categorized and analyzed.
\textit{Conclusion:} This mapping study provides the first systematic 
exploration of the state-of-art on software startup research. The existing 
body of knowledge is limited to a few high quality studies. Furthermore, the 
results indicate that software engineering work practices are chosen 
opportunistically, adapted and configured to provide value under the constrains 
imposed by the startup context. 
\end{abstract}

\begin{keyword}
%% keywords here, in the form: keyword \sep keyword
\small
Software Development \sep Startups \sep Systematic Mapping Study

%% MSC codes here, in the form: \MSC code \sep code
%% or \MSC[2008] code \sep code (2000 is the default)

\end{keyword}
\end{frontmatter}

%%
%% Start line numbering here if you want
%%
%\linenumbers

%% main text

\section{Introduction}  
\label{sect:intro} 
A wide body of knowledge has been created in recent years through several
empirical studies, investigating how companies leverage software engineering
(SE)~\cite{Kitchenham2009, da_silva_six_2011}. However, research on software 
development activities in newly created companies is scarce. In the past, very 
few publications have identified, characterized and mapped work practices in
software startups~\cite{Sutton2000} and no structured investigation of the area 
has been performed. Indeed, none of the systematic literature 
reviews~\cite{Kitchenham2007} or mapping studies~\cite{Budgen2007} in software 
engineering (see the tertiary review by Zhang and 
Babar~\cite{zhang_systematic_2013}) address the startup phenomenon.

Understanding how startups take advantage from work practices is 
essential to support the number of new businesses launched
everyday\footnote{According to a recent study, solely in the US
\textit{``startups create an average of 3 million new jobs 
annually''}~\cite{Formation2010}.}. 
New software ventures such as \textit{Facebook}, \textit{Linkedin}, 
\textit{Spotify}, \textit{Pinterest}, \textit{Instagram}, and \textit{Dropbox}, 
to name a few, are examples of startups that evolved into successful 
businesses. Startups typically aim to create high-tech and innovative products, 
and grow by aggressively expanding their business in highly scalable markets.

Despite many successful stories, self-destruction rather than 
competition drives the majority of startups into failure 
within two years from their creation~\cite{Crowne2002}. Software startups face 
intense time-pressure from the market and are exposed to tough competition, 
operating in a chaotic, rapidly evolving and uncertain 
context~\cite{Maccormack2001,Eisenhardt1998}. Choosing the right features to 
build and adapting quickly to new requests, while being constrained by limited 
resources, is crucial to the success in this environment~\cite{Sutton2000}.

From a software engineering perspective startups are unique, since they
develop software in a context where processes can hardly follow a prescriptive
methodology~\cite{Coleman2008}. Startups share some characteristics with other
contexts such as small companies and web engineering, and present a combination 
of different factors that make the development environment different from
established companies~\cite{Blank2005}. Therefore, research is needed to
support startups' engineering activities, guiding practitioners in taking 
decisions and avoiding choices that could easily lead business 
failure~\cite{Kajko-Mattsson2008,Coleman2005}.

The goal of this paper is to identify and understand the main contributions of 
the state-of-art towards software engineering in startups. To this end, we 
perform a systematic mapping study (SMS)~\cite{Budgen2007,Petersen2007} aimed 
at:
\begin{enumerate}[\textbullet]
  \item characterizing the state-of-art research on startups
  \item understanding the context that characterizes startups
  \item determining the potential for technology transfer of the state-of-art 
research on startups
  \item extracting and analyzing software development work practices used in 
startups
\end{enumerate}

The systematic map consists of 43 primary studies that were identified from 
an initial set of 1053 papers. Practitioners may take advantage of the 
213 identified software engineering work practices, while considering however 
the studies' respective rigor and relevance assessments. Furthermore, this first 
systematic exploration on software startups provides researchers with directions 
for future work.

The remainder of this paper is structured as follows: Section~\ref{sect:bg} 
describes background and related work; Section~\ref{sect:method} illustrates
the research methodology and discusses validity threats; 
Section~\ref{sect:schema} introduces the classification schema, developed from 
the gathered data; Section~\ref{sect:results} presents the results of the 
mapping study. The state-of-art of software  development in startups is discussed in 
Section~\ref{sect:discussion}, whereas in Section~\ref{sect:work_practices} the 
reported software development work practices are analyzed.
Section~\ref{sect:concl} concludes the paper, answering the posed research 
questions and providing an outlook for future work.

\section{Background and Related Work}
\label{sect:bg}

Modern entrepreneurship, born more than thirty years 
ago~\cite{storey1982entrepreneurship}, has been boosted by the advent of the
consumer Internet markets in the middle of the nineties and culminated with
the notorious dot-com bubble burst of 2000~\cite{Perkins:1999:IBI:555126}. 
Today, with the omnipresence of the Internet and mobile devices, we are
assisting to an impressive proliferation of software ventures - metaphorically 
referred as the startup bubble. In fact, easy access to potential markets and 
low cost of services distribution are appealing conditions for modern 
entrepreneurs~\cite{Marmer2011}. Inspired by success stories, a large
number of software businesses are created every day. However, the great
majority of these companies fail within two years from their 
creation~\cite{Crowne2002}.

\subsection{Software Startups}
\label{singularity}
An early account for the term \textit{startup} in the SE literature can be 
found in Carmel~\cite{Camel1994a} who studied in 1994 the time-to-completion in 
a young package firm. He noticed how these companies were particularly 
innovative and successful, advocating the need for more research on their 
software development practices so as to replicate success and try to transfer it 
to other technology sectors.

Sutton~\cite{Sutton2000} provides a characterization of software startups,
defined by the challenges they are faced with:

\begin{enumerate}[\textbullet]
\item little or no operating history - startups have little accumulated 
experience in development processes and organization management.
\item limited resources - startups typically focus on getting the product out, promoting the product and building up strategic alliances.
\item multiple influences - pressure from investors, customers, partners and 
competitors impact the decision-making in a company. Although 
individually important, overall they might be inconsistent.
\item dynamic technologies and markets - newness of software companies often require to develop or operate with disruptive technologies\footnote{A new technology that unexpectedly displaces an established technology. It does not rely on incremental improvements to an already established technology, but rather tackles radical technical change and innovation \cite{Christensen1997}.} to enter into a high-potential target market.
\end{enumerate}

One of the objectives of this SMS is to understand the context that 
characterizes startups and to what extent Sutton's definition from 2000 has been 
adopted or broadened.

% In this study we refer to software startups as newly created companies with no
% operating history and fast in producing cutting-edge technologies.
% These companies develop software under highly uncertain conditions, tackling
% fast-growing markets under severe lack of resources.
\subsection{Startup Lifecycle}
The lifetime of a startup company, from idea conception to the maturity level,
has been identified and reported from different perspectives (e.g. 
market~\cite{Blank2011a} and innovation \cite{Heitlager2006}). A prominent 
contribution, from a SE viewpoint, is the model presented by 
Crowne~\cite{Crowne2002}, who synthesized the startup lifecycle in four stages. 
The startup stage is the time when startups create and refine the idea 
conception, up to the first sale. This time frame is characterized most from 
the need to assemble a small executive team with the necessary skills to start 
to build the product. The stabilization phase begins from the first sale, and it 
lasts until the product is stable enough to be commissioned to a new customer 
without causing any overhead on product development. The growth phase begins 
with a stable product development process and lasts until market size, share and 
growth rate have been established. Finally, the startup evolves to a mature 
organization, where the product development becomes robust and predictable with 
proven processes for new product inventions.

\subsection{Software Development in Startups}
 \label{sect:bg:swdev}

The implementation of methodologies to structure and control the development
activities in startups is a major challenge for engineers~\cite{Coleman2008}.
In general, the management of software development is achieved through the
introduction of software processes, which can be defined as ``the coherent set
of policies, organizational structures, technologies, procedures, and
artifacts that are needed to conceive, develop, deploy and maintain a software
product''~\cite{Fuggetta:2000:SPR:336512.336521}. Several models have been
introduced to drive software development activities in startups, however
without achieving significant
benefits~\cite{Coleman2008a,Coleman2008,Sutton2000}.

In the startup context, software engineering (SE) faces complex and
multifaceted obstacles in understanding how to manage development processes.
Startups are creative and flexible in nature and reluctant to introduce
process or bureaucratic measures which may hinder their natural
attributes~\cite{Sutton2000,Bach1998}. Furthermore, startups have very limited
resources and typically wish to use them to support product development
instead of establishing processes~\cite{Coleman2008, Heitlager2007}. Some
attempts to tailor lightweight processes to startups reported basic failure
of their application~\cite{Martin2007}. Rejecting the notion of repeatable and
controlled processes, startups prominently take advantage of unpredictable,
reactive and low-precision engineering
practices~\cite{Sutton2000,Tanabian2005,Chorev2006,Kakati2003}.

Product-oriented practices help startups in having a flexible team, with 
workflows that leave them the ability to quickly change the direction according 
to the targeted market~\cite{Heitlager2007,Sutton2000}.
Therefore, many startups focus on team productivity, asserting more control to 
the employees instead of providing them rigid 
guidelines~\cite{Tanabian2005,Chorev2006, Kakati2003}.

Startups often develop applications to tackle a high-potential target market
rather than developing software for a specific
client~\cite{Marmer2011,ISI:000317949700032}. Issues related to this market
type are addressed in literature by market-driven software
development~\cite{Alves2006}. In this regard, researchers emphasize the
importance of time-to-market as a key strategic
objective~\cite{dagMDR,sawyer99}. In fact, startups usually operate in 
fast-moving and uncertain markets and need to cope with shortage of resources.
Other peculiar aspects influencing software development in the  market-driven
context are related to requirements, which are reported to be  often
``invented by the software company''~\cite{512553}, ``rarely
documented''~\cite{Karlsson02challengesin}, and can be validated only after
the product is released to market~\cite{dahl2003,Keil:1995}. Under these
circumstances, failure of product launches are largely due to ``products not
meeting customer needs''~\cite{Alves2006}.

%MUN_DELETE if we need to define a term, we should do it inline and add 
% references.
% \subsection{Terminology}
% Here we should define 
% \begin{enumerate}[\textbullet]
% \item "established companies"
% \item "software-intensive"
% \item "software development" and 
% \item differences between sw.dev in startups and established.  
% \item Mention web and small. 
% \item Engineering activities: the activities needed to bring a product from 
% idea to market. Traditional engineering activities are, among others, 
% requirements engineering, design, architecture, implementation,testing.
% \item "process" 
% \item work practices
% \item Software product: any software product and/or software service.
% \end{enumerate}

% Related work
% "Six years of systematic literature reviews in software engineering: An 
% updated tertiary study" --> 2008-2009
% Systematic literature reviews in software engineering – A systematic
% literature review" --> 2004-2007
% "Systematic literature reviews in software engineering – A tertiary study" 
% --> 2004-2008
% Google scholar: (software engineering) AND ("review of studies" OR "structured 
%review" OR "systematic review" OR "literature review" OR "literature analysis" 
%OR "in-depth survey" OR "literature survey" OR "meta analysis" OR "past 
%studies" OR "subject matter expert") --> 2010 - now

\subsection{Related work}
The prospects of evidence-based software 
engineering~\cite{kitchenham_evidence-based_2004} have motivated researchers to 
conduct systematic literature reviews and mapping studies. Zhang and 
Babar~\cite{zhang_systematic_2013} report on 148 SLR's and SMS's published 
between 2004 and 2010. However, none of these reviews nor the reviews conducted 
up to February 2014\footnote{We performed an automatic search with 
the search string published by Zhang and Babar~\cite{zhang_systematic_2013}.}, 
investigated software engineering in the context of startups.
Nevertheless, there exist reviews that studied software engineering topics 
pertinent to specific contexts or environments (as opposed to reviews that 
investigated individual software engineering technologies, e.g. feature 
location~\cite{dit_feature_2011} or search-based software 
testing~\cite{afzal_systematic_2009}) that we consider as related work.
Small and medium-sized enterprises (SMEs) and startups possibly share some 
characteristics, such as the low number of employees (fewer than 
250~\cite{_enterprise_2014}) and limited 
resources~\cite{Kamsties1998,Laporte2008}. Hence, reviews that study the 
literature on SMEs are relevant related work.

Pino et al.~\cite{pino_software_2008} studied the adoption of software process 
improvement approaches in SMEs~\cite{Richardson:2007:GEI:1262538.1262586}. 
They point out that very few of the SMEs that were part of the reviewed studies 
did achieve one of the pursued certifications, concluding that standard-driven, 
not tailored improvement initiatives are not suitable for small companies, 
confirming also Staples' et al. findings~\cite{Staples2007883}. Taticchi et 
al.~\cite{taticchi_performance_2010} observe a similar situation in the area of 
business performance measures and management (PMM). Their review identifies a 
lack of PMM models specifically adapted to SMEs, speculating that non-adoption 
stems from fear of costs and benefits incomprehension.

Thorpe et al.~\cite{IJMR:IJMR116} reviewed the literature on using knowledge 
within SMEs. Managers/entrepreneurs are an important organizational resource in 
SMEs as they are drivers for creating knowledge. This knowledge is best 
encoded in organized routines that allow widespread sharing within the firm. 
The challenge is to provide enough structure, allowing knowledge creation and 
sharing to scale, without limiting creativity and learning.

Rosenbusch et al.~\cite{Rosenbusch2011441} studied the 
innovation-performance relationship in SMEs by conducting a meta-analysis 
of 42 empirical studies that cover 21,270 firms. Interesting to our studied 
context is their finding that innovation has a stronger impact on younger firms 
than on more established SMEs. Furthermore, evidence suggests that for small and 
young firms it is more beneficial to conduct internal innovation projects 
than seeking innovation by collaborating with external partners. 

Common to these reviews, looking at different aspects of SMEs, is the 
recognition that properties of small firms require solutions and 
technologies that are adapted to that specific context. Similarly, we argue 
that startups, differing from SMEs in terms of their operating history, outside 
influences and market dynamism, require software development solutions adapted 
to their context. This SMS seeks to evaluate, synthesize and present the 
empirical findings on software development in startups to date and provide an 
overview of researched topics, findings, strength of evidence, and implications 
for research and practice.

\section{Research methodology}\label{sect:method}
\begin{figure*}
\centering
\includegraphics[scale=0.8]{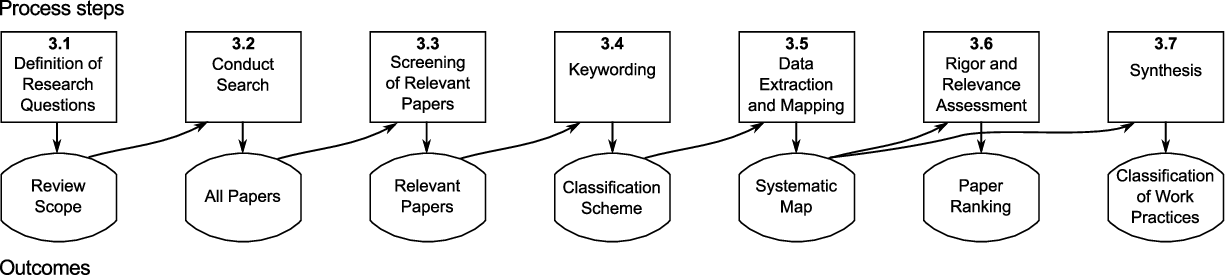}
\caption{Systematic mapping process (adapted from Petersen et 
al.~\cite{Petersen2007})}
\label{fig:ms:sms}
\end{figure*}
We chose to perform a systematic mapping study (SMS), which is capable of 
dealing with wide and poorly-defined areas~\cite{Petersen2007,Kitchenham2007}. A 
systematic literature review (SLR)~\cite{Kitchenham2007} would 
have been a less viable option due to the breadth of our overall research 
question: \emph{What is the state-of-art in literature pertaining to 
software engineering in startups?} 

The review in this paper follows the guidelines developed by 
Kitchenham and Charters~\cite{Kitchenham2007} and implements the 
systematic mapping process proposed by Petersen et al.~\cite{Petersen2007}. 
Figure~\ref{fig:ms:sms} illustrates the adopted process, whereas the individual 
steps are explained in Subsections~\ref{sub:rq}-~\ref{sub:synthesis}. 
Note that rigor and relevance assessment is an extension attributed to Ivarsson 
and Gorschek~\cite{Ivarsson2010} and synthesis is based on the constant 
comparison method proposed by Strauss and Corbin~\cite{QualitativeResearch}.

The SMS procedure was led by the first and second authors, who performed the 
steps in Figure~\ref{fig:ms:sms} in a co-located environment, i.e. working 
together on a single computer screen. Note-taking during screening of papers and 
keywording alleviated the resolution of conflicts among the reviewers during 
data extraction and rigor and relevance assessment. If disagreement persisted, 
an in-depth review of the paper was performed and, if necessary, the third and 
fourth authors were consulted to take a final decision.
%MUN_20131115: This explanation is not well formulated and does not contribute 
%to the understaning why you performed all steps together instead of splitting 
%up some work. Hence I delete it. Maybe the reviewers don't care....

%We decided not to split up during the some steps and perform the review
%together to progressively build shared criteria while improving our knowledge
%of nearby fields.

%NP OK

\subsection{Definition of Research Questions} % (fold)
\label{sub:rq}

The research question driving this mapping study is: \emph{What is the 
state-of-art in literature pertaining to software engineering in startups?} 
To answer this question, we state the following sub-questions:

\begin{enumerate}[\textbullet]
 \item RQ1 What is the context that characterizes software development in 
startups?
 \item RQ2 To what extent does the research on startups provide reliable and 
transferable results to industry?
 \item RQ3 What are the reported work practices in association with software 
engineering in startups?
\end{enumerate}

With RQ1 we intend to understand the properties that characterize the nature 
of software development in startups. Such a characterization illustrates the 
dimensions by which startups are defined in the state-of-art. With RQ2 we 
intend to determine the scientific evidence of the results reported in 
the state-of-art, allowing researchers to identify worthwhile avenues for 
further work and providing practitioners a tool to navigate within the 
state-of-art. With RQ3 we intend to identify the software engineering practices 
applied in startups, providing a basis for determining necessary further 
research.

\subsection{Conduct Search} % (fold)
\label{sub:research_process_overview}
We identified the primary studies by exercising a search string on scientific 
databases. The search string is structured according to population, 
intervention and comparison, as proposed by Kitchenham and 
Charters~\cite{Kitchenham2007}. We omitted however the outcome and context
facet from the search string structure as our research questions do not warrant 
a restriction of the results to a particular outcome (e.g. effective/not 
effective work practices) or context (e.g. startups in a specific product 
domain). 

Table~\ref{tab:ms:search-keywords} lists the final used keywords. The core 
concepts, representing population, intervention and comparison, are derived 
from our research questions. Following Rumsey's 
guidelines~\cite{how-find-information}, we identified synonymous, 
related/broader/wider concepts, alternative spelling and part of speech for each 
core concept. Note that we did not include specific keywords from existing 
startup definitions (e.g. Sutton~\cite{Sutton2000}, discussed in 
Section~\ref{singularity}) to the population set of terms as this could 
have biased the search.

%Table of keywords  ---- ---- ---
\begin{table}
\caption{Population, intervention and comparison search string keywords}
\label{tab:ms:search-keywords}
%\centering
\footnotesize
\begin{tabular}{p{0.8in}p{2.3in}}
\toprule
\multicolumn{1}{c}{Core concepts} &  \multicolumn{1}{c}{Terms} \\
\midrule
\midrule
Software Startups & software startup*; software start-up*; 
early-stage firm*; early-stage compan*; high-tech venture*; high-tech 
start-up*; start-up compan*; startup compan*; lean startup*; lean start-up*;  
software package start-up*; software package startup*; IT start-up*; IT 
startup*; software product startup*; software start up*; internet start-up*; 
internet startup*; web startup*; web start-up*; mobile startup*; mobile 
start-up*;  \\ \midrule
Development & develop*; engineer*; model*; construct*; 
implement*; cod*; creat*; build*; \\ \midrule
Strategy & product*; service*; process*; methodolog*; tool*; 
method*; practice*; artifact*;  artefact*; qualit*; ilit*; strateg*; software; 
\\
\bottomrule
\end{tabular}
\end{table}
%endtable

The selected scientific databases on which we performed the search are 
shown in Table~\ref{tab:ms:databases}, along with the number of publications 
retrieved from each database (up to December 2013). 
We selected the databases considering their coverage and use in the 
domain of software engineering, and their ability to handle advanced queries, 
following the example of Barney et al.~\cite{Barney2012}.

To increase publication coverage we also used Google Scholar, which 
indexes a large set of data, both peer and non-peer reviewed. Then, we proceeded 
to the customization of the search string, adapting the syntax to the 
particular database\footnote{The individual search strings are available in 
the supplementary material~\cite{paternoster_supplementary_2013}.}.
%table retrieved overview  ---- ---- ---
\begin{table}[H]
\caption{Selected databases and retrieved papers}
\label{tab:ms:databases}
\footnotesize 
\centering
\begin{tabular}{llc}
\toprule
ID & \multicolumn{1}{c}{Database} & Papers   \\
\midrule
\midrule
  A & Inspec/Compendex (www.engineeringvillage2.org) & 640 \\
  B & IEEE Xplore (ieeexplore.ieee.org) & 132 \\
  C & Scopus (www.scopus.com) & 468 \\
  D & ISI Web of Science (wokinfo.com) & 293 \\
  E & ACM Digital Library (dl.acm.org/advsearch.cfm) & 78 \\
  F & Google Scholar (scholar.google.com) & 158\\
\midrule
      & \textbf{Total} & \textbf{1769} \\
\bottomrule
\end{tabular}
\end{table}
%endtable
\subsection{Screening of Relevant Papers} % (fold)
\label{sub:retr_screen}
The criterion that guided the inclusion of a paper was that the study 
presented a contribution to the body of knowledge on software development in 
startups. A contribution can be in the form of an experience report, applied 
engineering practices, development models or lessons learned.
We excluded search results that were:
\begin{enumerate}[\textbullet]
\item not peer-reviewed (grey literature, books, presentations, blog posts, 
etc.)
\item not written in English
\item clearly obsolete (more than 20 years old)
\item related to non-software companies (biotech, manufacturing, electronics, 
etc.)
\item related to established companies (VSE, SME, research spin-offs)
\item related to technicalities of startups (algorithms, programming 
languages, etc.)
\end{enumerate}

\begin{figure}[H]
\centering
\includegraphics[width=0.5\textwidth,keepaspectratio=true]{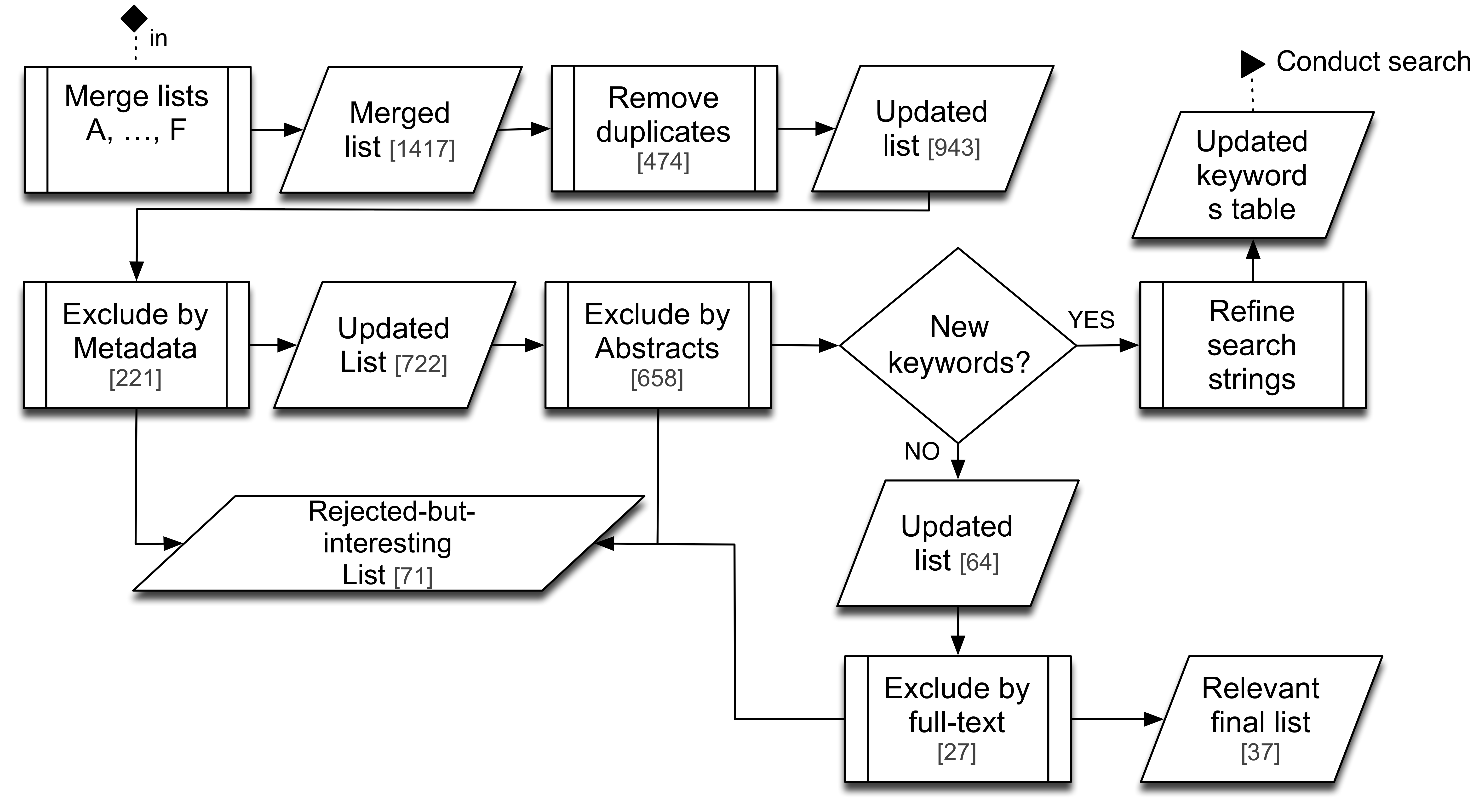}
\caption{Screening of papers workflow}
\label{fig:ms:screening}
\end{figure}

For the screening of papers we followed the workflow in 
Figure~\ref{fig:ms:screening}.

%%startFigure
%%end figure

With the support of a reference management tool~\cite{bibdesk} we merged 
the six result lists from the search in the scientific databases. Then, we 
removed duplicated items in two steps: first we used the reference management 
tool to automatically detect duplicates based on meta-data (author, publication 
year and title). Then, we manually deleted instances that were not detected as 
duplicates by the tool, resulting in 1053 papers.

Then we analyzed the metadata (title, keywords, publication year and type) 
to identify papers that matched the exclusion criteria, resulting in 722 papers.
In a more in-depth review, we analyzed the abstract of each paper, determining 
whether it matched our inclusion criterion, resulting in 64 papers. As 
indicated in Figure~\ref{fig:ms:screening}, we improved the search strings 
while reading the abstracts, adding new keywords identified in retrieved 
papers and iteratively conducting a new search.

In case of a disagreement among the reviewers or incomplete abstracts we read 
the entire paper, leading eventually to the final set of 43 primary studies. 
During the screening process we kept track of the rationale for each exclusion,
as shown in Table~\ref{tab:ms:exclusion_rationale}.

\begin{table}[H]
\caption{Rationale for excluded papers}
\label{tab:ms:exclusion_rationale}
\footnotesize 
\centering
\begin{tabular}{ll}
\toprule
\multicolumn{1}{c}{Rationale} & \multicolumn{1}{c}{Amount}   \\
\midrule
\midrule
Duplicate & 474\\
Non-software industry & 409 \\
Related to software startups but not SE perspective & 259 \\
Not software/startup related & 122 \\
Related to software but not startups & 85 \\
Academic settings & 70\\
Not peer-reviewed & 41 \\
Full-text non available & 20 \\
Outdated & 4 \\
Not in English language & 4 \\
\midrule 
Total retrieved & 1531 \\
Total excluded & 1488  \\
\textbf{Total included} & \textbf{43} \\
\bottomrule
\end{tabular}
\end{table}

% subsection retr_screen (end)

\subsection{Keywording} % (fold)
\begin{figure}[H]
\centering
\includegraphics[width=0.5\textwidth,keepaspectratio=true]{%
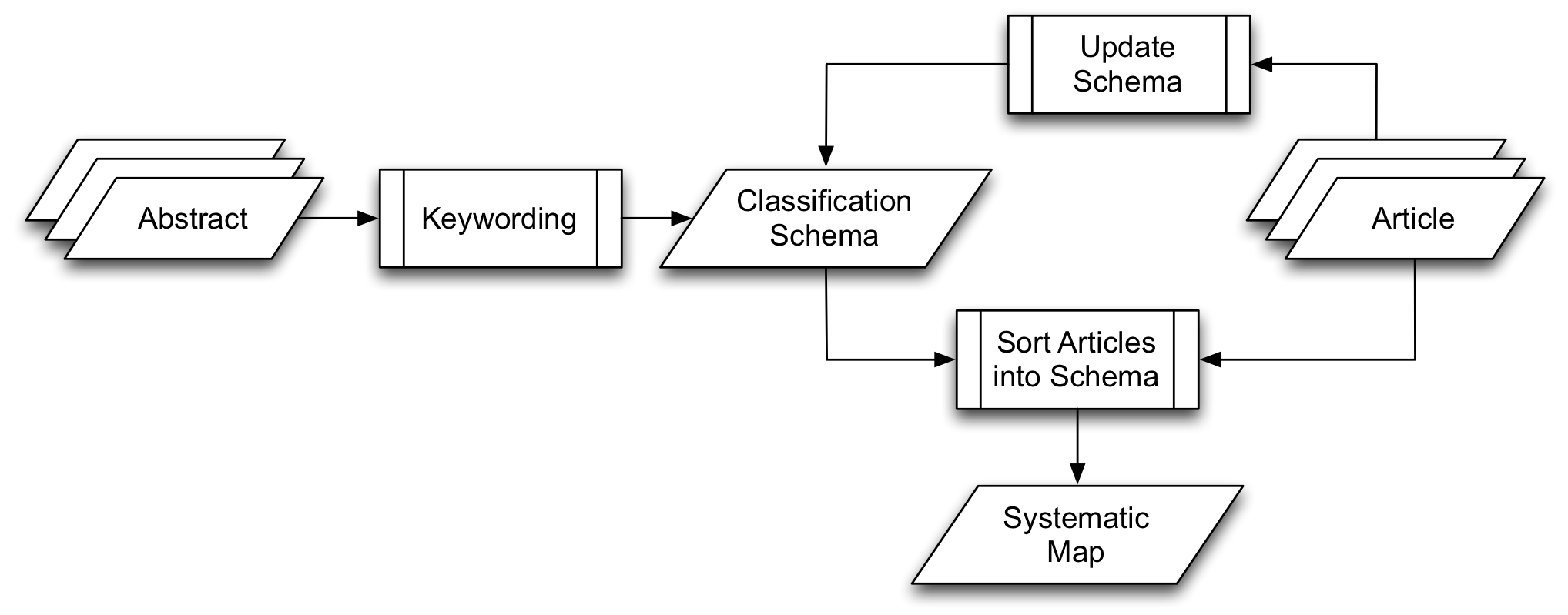}
\caption{Workflow for classification schema creation (adapted 
from~\cite{Petersen2007})}
\label{fig:ms:keywording}
\end{figure}

\label{sub:keywording}
The goal of keywording is to efficiently create a classification schema, 
ensuring that all relevant papers are taken into account~\cite{Petersen2007}. 
We followed the process illustrated in Figure~\ref{fig:ms:keywording}.

The first step consisted in reading the abstracts of the primary studies,
assigning them a set of keywords to identify the main contribution area of the
paper. Then we combined the keywords forming a high-level set of categories, 
leading to a rough understanding of the research area represented by the
primary studies. By progressively fitting the papers into categories, the 
schema underwent a refinement process, being continuously updated to account for 
new data. When performing data extraction and mapping 
(Subsection~\ref{sub:data_extraction}), we annotated the classification with 
evidence from the respective paper, further refining the schema and sorting.
The resulting classification schema is discussed in Section~\ref{sect:schema} 
and used in the analysis of the results in 
Sections~\ref{sect:discussion} and~\ref{sect:work_practices}.

% subsection mapping (end)

\subsection{Data Extraction and Mapping} % (fold)
\label{sub:data_extraction}

After we defined the classification schema, resulting from the keywording 
process, we proceeded to systematically extract data from the
primary studies. For each paper, we filled a spreadsheet, sorting it 
into the classification schema and extracting the following data, inspired by 
other similar studies~\cite{Dyba2008,Shepperd2007}:
\begin{enumerate}[\textbullet]
\item Article title
\item First author
\item Year of publication
\item Synthesis of results (one-line)
\item Keywords
\end{enumerate}

We took advantage of the data extraction process to identify an additional
relevant aspect which emerged while reading abstracts and the full text: the
recurrent patterns of common attributes among startup companies resulted in 
themes that are reported in 
Subsection~\ref{sub:contextual_features_of_startups}. Moreover, we screened the 
bibliography of each paper, identifying other
possible relevant studies to our research, adopting the snowballing
technique\footnote{Note that we didn't identify any additional papers. In case 
more relevant papers would have been retrieved, a re-iteration of the 
keywording step would have been necessary (see
Subsection~\ref{sub:keywording}).}~\cite{snowballing}.

% subsection data_extraction (end)

\subsection{Rigor and relevance assessment} % (fold)
\label{sub:rigor_and_relevance}

A major challenge of SE is to transfer research results and knowledge to
practitioners, showing the findings' validity and concrete 
advantages~\cite{Ivarsson2010}. To assess how results are presented in the 
primary studies, we extended the traditional SMS framework with an additional 
step, that is, the evaluation of the papers' rigor and relevance 
(see Figure~\ref{fig:ms:sms}). With this extension we compensate for the SMS' 
limitation of not assessing the quality of the mapped studies by developing and 
using a simple ranking function.

We use a systematic and validated model~\cite{Ivarsson2010} to evaluate the 
scientific rigor and the industrial relevance of each primary study. The model 
provides a set of rubrics to measure 
rigor and relevance, dividing these two factors into different aspects, and 
quantifying the extent to which each aspect is considered
in the study (see Ivarson and Gorschek~\cite{IvarssonGorsREj} for an 
application of the model).

Rigor refers to the precision or exactness of the used
research method and how the study is presented. We considered aspects relating 
to:
\begin{enumerate}[\textbullet]
\item Context - description of development mode, speed, company maturity and
any other important aspects where the evaluation is performed.
\item Study design - description of the measured variables, treatments, 
used controls and any other design aspects.
\item Validity - description of different types of validity threats.
\end{enumerate}

Relevance refers to the realism of the environment where the study is performed 
and the degree to which the chosen research method contributes to the 
potential of transferring the results to practitioners. We considered aspects 
relating to:
\begin{enumerate}[\textbullet]
\item Subjects - use of subjects who are representative of the intended users
of the technology.
\item Context - use of settings representative of the intended usage setting.
\item Scale - use of a realistic size of the applications.
\item Research method - use of a research method that facilitates
investigating real situations and relevant for practitioners.
\end{enumerate}

Aspects related to the rigor of the study are scored at three levels: weak
(0), medium (0.5) and strong (1). Aspects related to relevance are scored 1 if
contributing, 0 otherwise. The detailed rubrics, used to evaluate the studies,
can be found in Ivarsson and Gorschek~\cite{Ivarsson2010}. To obtain the
study's final score, we sum the individual scores of the rigor and relevance
aspects.

In order to rank the papers, we defined a function incorporating the 
classification schema, rigor and relevance scores, and two additional factors 
that characterize the publication type and year.
The ranking function provides a rough estimation of the value that a 
paper provides to practitioners and the research community, giving a 
stronger weight to recent rigorous journal publications entirely devoted to the 
topic and presenting empirical results relevant to practitioners. 
We used tables for converting each factor into an arbitrary numerical value in 
the range between 0 and 10. The conversion tables used to quantify the
internal score of each factor are shown in the supplementary 
material~\cite{paternoster_supplementary_2013}, while the limitations of this 
approach are discussed in Subsection \ref{sect:rm:ms:validity:sele}. The final 
ranking of the 43 primary studies is discussed in 
Subsection~\ref{sub:transferability}.

\subsection{Synthesis} 
 \label{sub:synthesis}
In the synthesis we identified the main concepts from each primary study, using 
the original author's terms in a one line statement. Those main concepts were then
organized in tabular form to enable comparisons across studies and translation 
of findings into higher-order working practices and classification categories. 
We used the classification categories from Section~\ref{sect:schema}. This 
process is analogous to the method of constant comparison used in qualitative 
data analysis~\cite{QualitativeResearch}.

In Section~\ref{sect:work_practices} we present the identified work practices, 
discussing their application in the startup context, their benefits and 
liabilities, and putting them in perspective with the results of other 
studies. In summary, synthesis is achieved by:

\begin{enumerate}[\textbullet]
 \item Identification of a set of working practices and relative
classification categories.
 \item Documentation of advantages and disadvantages of reported results.
 \item Elaboration of gaps on the applicability of working practices in 
startup contexts.
\end{enumerate}

\subsection{Threats to validity} 
 \label{sub:validity}

We identified potential threats to the validity of the systematic mapping and
its results, together with selected mitigation strategies. The structure of this
Section follows Unterkalmsteiner et al.~\cite{Unterkalmsteiner2012}.

%--------------------------%%--------------------------%%--------------------------%%--------------------------%%--------------------------%%--------------------------%%--------------------------%
\subsubsection{Publication bias}
\label{sect:rm:ms:validity:pub}

Systematic reviews suffer from the common bias that positive outcomes are more
likely to be published than negative ones~\cite{Kitchenham2007}. This 
can be observed also in our mapping study which includes few papers on failed 
startup endeavors and studies, while the success rates of startups is generally 
rather low. It is unlikely that research is performed only in collaboration 
with successful startups. Nevertheless, we do not consider this as a major 
threat as this bias is orthogonal to our study aim, mapping the state-of-art on 
startup research. Still, this bias takes away some of the possibilities to 
analyze reported work practices with regard to their performance.

%MUN_20131115: The argumentation below is not related to the publication bias 
% but to the aspect that we are mapping state-of-the-ART and not 
% state-of-PRACTICE. Hence your argumentation is not valid. Removed and 
% replaced by my own interpretation. 

%NP OK

%Moreover some  of the companies considered in these study have
%been not only passively observed, but actively pushed to adopt some specific
%work practice by the researcher for the sake of the study.  Some of these
%practices wouldn't have otherwise emerged naturally in the context of a 
%startup.
%Therefore the observation extracted about work practices in startups might not
%reflect completely the industry situation but some particular instance of it.

%--------------------------%%--------------------------%%--------------------------%%--------------------------%%--------------------------%%--------------------------%%--------------------------%
\subsubsection{Identification of primary studies}
\label{sect:rm:ms:validity:thre}

The approach we used to construct the search string (see 
Subsection~\ref{sub:research_process_overview}) aimed to be inclusive with respect 
to the number of retrieved papers, related to software development in startups.

However, a limitation of the current search string lies in the exclusion
of the stand-alone terms ``startup'' and ``start-up''. These individual terms 
lead to unmanageable search results (more than 20000 papers) that are 
mostly irrelevant as they are related to the English phrasal verb ``to start 
up'', largely used in many disciplines to indicate the commencing moments of an 
engine. Therefore, to mitigate the risk of excluding potential relevant primary 
studies, we constructed a search string containing qualifiers to the term 
``startup'', e.g. ``IT startup'', and included synonyms, validating our search strings with the
support of librarians specialized in software engineering.

Still, the precision (ratio of retrieved relevant and all retrieved 
papers~\cite{Saracevic1995}) of the used search string is low (43 out of 
1053, 4\%). However we were not interested in obtaining high
precision as much as we aimed to obtain a high recall (ratio of existing relevant 
papers~\cite{Saracevic1995}).  The risk of 
excluding relevant primary studies is further mitigated by the use of multiple 
databases, which cover the majority of scientific publications in the field.

We were not able to retrieve 20 papers since they were neither available in
online catalogs, in the three libraries we consulted, nor on request from the 
authors. However, this is a minor risk as we had access to their titles, 
keywords and abstracts, which gave us a good degree of confidence that they 
were not relevant. Additionally, considering our 4\% precision rate, the 
number of potentially relevant primary studies would be less than 1.

We noticed high inconsistency in the use of the term ``startup''
by different researchers, even in the same area. For example, 
Sutton~\cite{Sutton2000} distinguishes startups from established companies by
characterizing startups according to their extreme lack of resource, newness 
and immaturity (in a process sense). On the other hand, Deias et 
al.~\cite{Deias} define their company as a startup, with more then 150 
employees and resources available to certify the quality of their development 
process. Under these conditions, the attempt to identify a body of knowledge 
and research scope has been highly challenging. Therefore, we had to 
identify and analyze multiple and conflicting definitions. 

Moreover, several contextual factors, not thoroughly analyzed in this study,
can be identified as important. Factors regarding the application domain or 
the market type could influence the adoption of working practices and
processes. However, in this study we compromised details regarding specific
context challenges in favor of a general overview of practices reported by
primary studies.

Finally, since startups and entrepreneurship in general are appealing for many 
sectors of the economy, an additional threat lies in the fact that some relevant
information can be found in other research areas, such as business innovation 
and marketing, not considered in this study.

\begin{table}
\caption{Classification schema}\label{tab:classificationschema}
\centering
\subfloat[Research type facet (adapted from 
Wieringa~\cite{Wieringa:2005})\label{subtab:researchtype}] {
\footnotesize
\begin{tabularx}{\linewidth}{ p{0.7in} X }
\toprule
Category & Description \\
\midrule
\midrule
Evaluation Research & The methodology is implemented in practice and an 
evaluation
of it is conducted. That means, it is shown how the research is implemented
(solution implementation) and what are the consequences of the
implementation in terms of benefits and drawbacks (implementation evaluation).
This also includes problems identified in industry. \\
Solution Proposal & A solution for a problem is proposed. The solution can be
either novel or a significant extension of an existing methodology. The
potential benefits and the applicability of the solution is shown by a small
example or a good line of argumentation.  \\
Philosophical Papers & These papers sketch a new way of looking at existing
things by structuring the field in form of a taxonomy or conceptual framework.
\\
Opinion Papers & These papers express the personal opinion of somebody whether
a certain technique is good or bad, or how things should have been done. They do
not rely on related work and research methodology. \\
Experience Papers & Experience papers explain what and how something has
been done in practice. It has to be the personal experience of the author. \\

\bottomrule
 \end{tabularx}
}\\
\subfloat[Contribution facet (adapted from 
Shaw~\cite{Shaw2003})\label{subtab:contribution}] {
\footnotesize
\begin{tabularx}{\linewidth}{ p{0.7in} X }
\toprule
Category &	Description\\
\midrule
\midrule
Model & Representation of an observed reality by concepts or related concepts 
after a conceptualization process. \\  
Theory & Construct of cause-effect relationships of determined results.  \\ 
Framework / Methods & Models related to constructing software or managing 
development processes. \\  
Guidelines & List of advises, synthesis of the obtained research results.  \\  
Lesson learned & Set of outcomes, directly analyzed from the obtained research 
results. \\   
Advice / Implications & Discursive and generic recommendation, deemed from 
personal opinions. \\    
Tool  & Technology, program or application used to create, debug, maintain or 
support development processes. \\
\bottomrule
 \end{tabularx}
}\\
\subfloat[Focus facet\label{subtab:focus}] {
\footnotesize
\begin{tabularx}{\linewidth}{p{0.7in} X }
\toprule
  Category & Description \\
\midrule
\midrule
Software development & Engineering activities used to write and maintaining the 
source code. \\ 
Process management & Engineering methods and techniques used to manage the 
development activities. \\
Tools and technology & Instruments used to create, debug, maintain and support
development activities. \\
Managerial / Organizational & Aspects that are related to software development,
by means of resource management and organizational structure. \\
\bottomrule
 \end{tabularx}
}\\
\subfloat[Pertinence facet\label{subtab:pertinence}] {
\footnotesize
\begin{tabularx}{\linewidth}{ p{0.3in} X }
\toprule
  Category &	Description\\
\midrule
\midrule
Full  & Entirely related (main focus) to engineering activities in software
startups.   \\
Partial & Partially related to  engineering activities in software startups.
Main research focus related to engineering activities. \\
Marginal &  Marginally related to engineering activities in software startups.
Main research focus different from engineering activities. \\
\bottomrule 
 \end{tabularx}
}
\end{table}

%--------------------------%%--------------------------%%--------------------------%%--------------------------%%--------------------------%%--------------------------%%--------------------------%
\subsubsection{Study selection and data extraction}
\label{sect:rm:ms:validity:sele}

Threats to study selection and data extraction~\cite{Dyba2008} have been 
mitigated with an up-front definition of the inclusion/exclusion 
criteria~\cite{Kitchenham2007}.
The selection of relevant primary studies can be further biased by personal
opinions of researchers executing the process. To mitigate this threat, we 
defined and documented a rigid protocol for the study selection and, by 
conducting the selection together and dedicating a reasonable amount of time to 
review conflicts, mutually adjusting each others' biases, as suggested by 
Kitchenham and Charters~\cite{Kitchenham2007}. 
%NP following paragraph needs review
%MUN_20131115: Moved from "Identification of primrary studies" to "Study 
%selection" threat, reviewed and reformulated...
%Since we did not  split-up the review process, another threat to the 
%validity of our study is the potential predominance of the opinion of one 
%author on the others. To mitigate this threat we had the possibility of having 
%the third author reviewing those inclusions where it was not possible to reach 
%an agreement amonth the first two authors.  However this condition never 
%actually realized.
The screening process is threatened by a potential predominance of the 
opinion of one reviewer over the other, since the first two authors performed 
the screening process collaboratively at one computer. This threat was 
mitigated by explicitly recording the exclusion rationale for each paper, 
requiring clear evidence from the paper to support the decision, and supporting 
the consensus creating process by consulting the history of previously taken 
decisions.

Another threat is related to researchers’ personal judgments, which can
interfere with the evaluation of rigor and relevance of selected studies.
Even though the rigor and relevance model provides guidelines and detailed 
rubric tables to support objective decisions, the evaluation depends on the 
reporting quality and not on the intrinsic quality of the study 
itself~\cite{Ivarsson2010}. As such, we used the scores only to rank but not to 
exclude studies from the selection. The ranking itself gives an indication of 
the study quality, the individual contribution needs however to be qualified by 
the reader~\cite{Ivarsson2010}.

The validity of the ranking function (see 
Subsection~\ref{sub:rigor_and_relevance}) is threatened by the arbitrarily chosen 
scores for each category and weights for each dimension.
To mitigate this threat, we used an automatic spreadsheet to compute the final 
scores, allowing us to adjust scores and weights, observing the effect of 
the final ranking in real time. For validating our ranking, we tried to modify
scores/weights values several times, and we observed that the final ranking
was not significantly altered by numerical adjustments, as long as we kept the 
ordering of concepts stable. 

 % subsection subsection_name (end)

\section{Classification schema} %-------- ## ----------------- ## ----------------- ## ----------------- ## ----------------- ## ---------
 \label{sect:schema}

In this section we present the classification schema that is adapted from other 
existing taxonomies or emerged from the keywording process. The schema consists 
of four facets:

\begin{enumerate}[\textbullet]
\item Research type: to represent the type of the undertaken study
\item Contribution type: to map the different types of the study outcomes
\item Focus: to describe the main focus of the research
\item Pertinence: to distinguish between studies entirely devoted to
engineering activities in startups and the ones that have a broader perspective
\end{enumerate}

The research type facet (Table~\ref{subtab:researchtype}) is used to 
distinguish between different types of studies, abstracting from the specific 
underlying research methodology. The research types were adapted from 
Wieringa~\cite{Wieringa:2005}.

The contribution facet (Table~\ref{subtab:contribution}), similarly to the 
taxonomy used by Shaw~\cite{Shaw2003}, describes the kind of contribution a 
study provides. Contribution types can be divided into weak (advises and 
implications, lessons learned, tools and guidelines) and 
strong (theory, framework/method and model) contributions.

The categories in the focus facet (Table~\ref{subtab:focus}) were obtained by 
clustering the sets of keywords identified in the keywording process 
(Subsection~\ref{sub:keywording}) and abstracting them to four categories.
We separated thereby studies concerning software development practices 
(e.g. writing user stories~\cite{Kajko-Mattsson2008}) from studies focused on 
higher-level process management (e.g. use Scrum 
methodology~\cite{Kuvinka2011}). Furthermore we distinguish
between studies focused on specific tools and technologies (e.g. use of post-it 
notes~\cite{Silva2005}) and work focused on managerial/organizational aspects 
in startups (e.g. operate in cross-functional settings~\cite{Camel1994a}).

The pertinence facet (Table~\ref{subtab:pertinence}) distinguishes the levels 
(full, partial, marginal) on which the study's research focus is  directed 
towards engineering activities in startups.

The classification schema in Table~\ref{tab:classificationschema} forms the 
basis for the systematic maps presented and discussed in 
Section~\ref{sect:results}.

\section{Results} %-------- ## ----------------- ## ----------------- ## ----------------- ## ----------------- ## ---------
 \label{sect:results}
This section presents the results of the systematic mapping study. From an 
initial sample of 1053 papers, we identified 43 primary studies
answering our research questions.

\subsection{Startup research categorization} % (fold)
\label{sub:publication_distribution}

\begin{figure}
\centering
\includegraphics[width=0.5\textwidth,keepaspectratio=true]{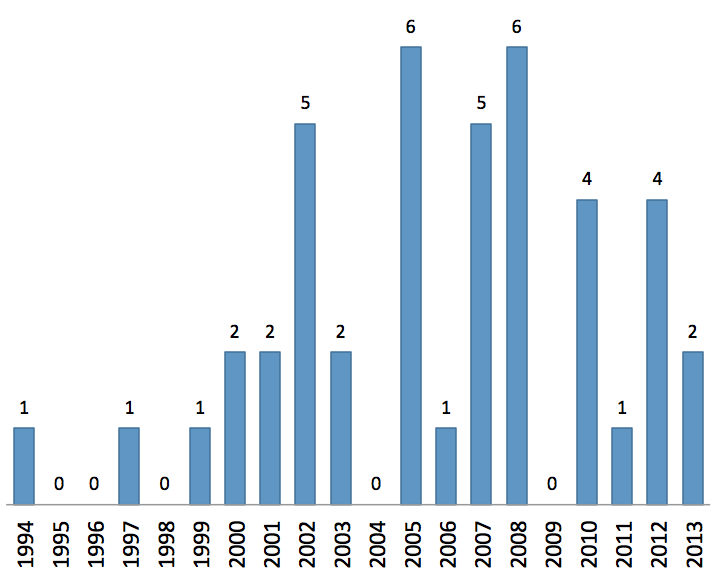}
\caption{Publication distribution-year}
\label{fig:ms:year}
\end{figure}

Figure~\ref{fig:ms:year} shows the publication years' frequency 
distribution, from 1994 to 2013.

%%startFigure

\begin{table*}[t]
\caption{ Systematic map overview }\label{tab:ms:refmap}
\centering
\footnotesize
\begin{tabular}{p{1.1in}p{1.1in}p{1.1in}p{1.3in}p{0.5in}}
    \toprule
     1st Author (year) & Research Type & Contribution & Focus & Pertinence 
\\
    \midrule
    \midrule
Coleman (2008) \cite{Coleman2008} & Evaluation Research & Model & Process 
Management & Full  \\
Kajko (2008) \cite{Kajko-Mattsson2008} & Evaluation Research & Model& Process 
Management & Full \\
Mater (2000) \cite{Mater2000} & Evaluation Research & Model & Managerial \& 
Organizational & Partial \\
H\"{a}sel  (2010) \cite{Hasel2010} & Evaluation Research & Model & Managerial 
\& Organizational & Marginal \\
Hanna (2010) \cite{Hanna2010} & Evaluation Research & Model & Managerial \& 
Organizational & Marginal \\
Chorev (2006) \cite{Chorev2006} & Evaluation Research & Model & Managerial \& 
Organizational & Marginal \\
Kakati (2003) \cite{Kakati2003}  & Evaluation Research & Model & Managerial \& 
Organizational & Marginal \\
Kim (2005) \cite{Kim2005} & Evaluation Research & Model & Managerial \& 
Organizational & Marginal \\
Coleman (2007) \cite{Coleman2007} & Evaluation Research & Theory & Process 
Management & Full  \\
Coleman (2008) \cite{Coleman2008a} & Evaluation Research & Theory& Process 
Management & Full  \\
Bosch (2013) \cite{bosch2013early}  & Evaluation Research & Framework \& Methods 
& Process Management & Full  \\
Midler (2008) \cite{Midler2008} & Evaluation Research & Framework \& Methods & 
Managerial \& Organizational & Partial \\
Yogendra (2002) \cite{Yogendra2002} & Evaluation Research & Guidelines & 
Managerial \& Organizational & Partial \\
Yoffie (1999) \cite{Yoffie1999} & Evaluation Research & Guidelines & Managerial 
\& Organizational & Marginal \\
Camel (1994) \cite{Camel1994a} & Evaluation Research & Lesson Learned & Software 
Development & Full  \\
Silva (2005) \cite{Silva2005} & Evaluation Research & Lesson Learned & Software 
Development & Full  \\
Jansen (2008) \cite{Jansen2008} & Evaluation Research & Lesson Learned & 
Software Development & Partial \\
Steenhuis (2008) \cite{Steenhuis2008} & Evaluation Research & Lesson Learned & 
Managerial \& Organizational & Marginal \\
Lai (2010) \cite{Lai2010} & Evaluation Research & Lesson Learned & 
Managerial \& Organizational & Marginal \\
Tingling (2007) \cite{Tingling2007} & Evaluation Research & Advice \& 
Implications & Software Development & Full  \\
Li (2007) \cite{Su-Chan2007} & Evaluation Research & Advice \& 
Implications & Software Development & Marginal \\
Blank (2013) \cite{ISI:000317949700032} & Solution Proposal & Framework \& 
Methods & Process Management & Partial \\
Zettel (2001) \cite{Zettel2001} & Solution Proposal & Framework \& Methods  & 
Software Development & Full  \\
Crowne (2002) \cite{Crowne2002} & Solution Proposal & Advice \& Implications & 
Software Development & Full  \\
Mirel (2000) \cite{Mirel2000} & Solution Proposal & Advice \& Implications & 
Managerial \& Organizational & Partial \\
Stanfill (2007) \cite{Stanfill2007} & Solution Proposal & Advice \& Implications 
& Managerial \& Organizational & Marginal \\
Himola (2003) \cite{Hilmola2003} & Solution Proposal & Advice \& Implications & 
Managerial \& Organizational & Marginal \\
Heitlager (2007) \cite{Heitlager2007} & Solution Proposal & Tool & Process 
Management & Full  \\
Yoo (2012) \cite{ISI:000312482500005} & Philosophical Paper & Model  & 
Managerial \& Organizational & Marginal \\
Yu (2012) \cite{20124315605867} & Philosophical Paper & Guidelines  & Managerial 
\& Organizational & Marginal \\
Fayad (1997) \cite{Fayad1997} & Philosophical Paper & Advice \& Implications & 
Process Management & Marginal \\
Bean (2005) \cite{Bean2005} & Philosophical Paper & Advice \& Implications & 
Tools \& Technology & Marginal \\
Sutton (2000) \cite{Sutton2000} & Opinion Paper & Advice \& Implications & 
Process Management & Full  \\
Tanabian (2005) \cite{Tanabian2005} & Opinion Paper & Advice \& Implications & 
Managerial \& Organizational & Marginal \\
Deakins (2005) \cite{Deakins2005} & Experience Paper & Model & Managerial \& 
Organizational & Partial \\
Ambler (2002) \cite{Ambler2002} & Experience Paper & Lesson Learned & Software 
Development & Full  \\
May (2012) \cite{6298105}  & Experience Paper & Advice \& Implications  & 
Software Development & Full  \\
Taipale (2010) \cite{Taipale2010} & Experience Paper & Advice \& Implications & 
Software Development & Full  \\
Deias  (2002) \cite{Deias} & Experience Paper & Advice \& Implications & 
Software Development & Full  \\
Wood (2005) \cite{Wood2005} & Experience Paper & Advice \& Implications & 
Software Development & Partial \\
Wall (2001) \cite{Wall2001} & Experience Paper & Advice \& Implications & 
Software Development & Partial \\
Kuvinka (2011) \cite{Kuvinka2011} & Experience Paper & Advice \& Implications & 
Software Development & Partial \\
Clark (2012) \cite{20124415623309}  & Experience Paper & Advice \& Implications 
& Process Management & Marginal \\
\bottomrule
\end{tabular}
\end{table*}

To characterize the main themes covered within the area of engineering 
activities in software startups, we used the set of keywords extracted from the 
study's abstract and author-defined keywords (when available in the metadata). 
From the 43 primary studies we 
extracted a total of 346 keywords (125 unique) averaging on about 8 keywords per 
paper. These formed the basis\footnote{The raw data is provided in the 
supplementary material~\cite{paternoster_supplementary_2013}.} for the focus 
and pertinence facet of the classification schema (Section~\ref{sect:schema}). 
Table~\ref{tab:ms:refmap} applies the classification schema on the primary 
studies, providing an overview of the field of startup research.

%MUN3 Deleted as does not at much to the results.
%To provide a visual representation of the studies' content, 
%Figure~\ref{fig:ms:cloud} illustrates the frequency of occurrence of each 
%keyword in a tag cloud\footnote{The tag cloud was generated with the online 
%service (http://tagcrowd.com/).}. The table containing the raw data is 
%provided 
%in the supplementary material~\cite{paternoster_supplementary_2013}.

%%startFigure
%\begin{figure}[H]
%\centering
%\includegraphics[width=0.5\textwidth,keepaspectratio=true]{figures/cloud.png}
%\caption{Keywords cloud overview}
%\label{fig:ms:cloud}
%\end{figure}
%%end figure

\begin{figure*}
\centering
\includegraphics[width=0.9\textwidth,keepaspectratio=true]{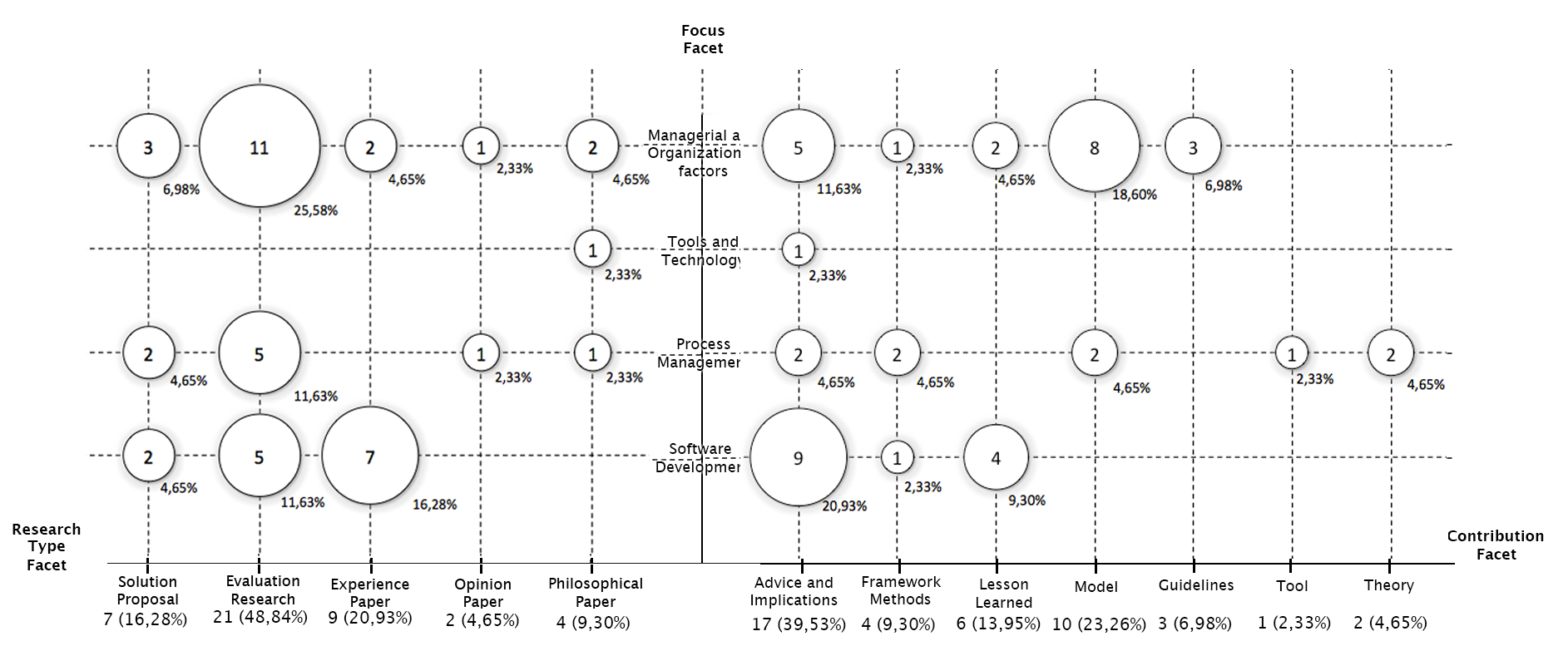}
\caption{Systematic map - \textit{Focus}, \textit{contribution} and \textit{research type}}
\label{fig:ms:fcr}
\end{figure*}

\begin{figure*}
\centering
\includegraphics[width=0.9\textwidth,keepaspectratio=true]{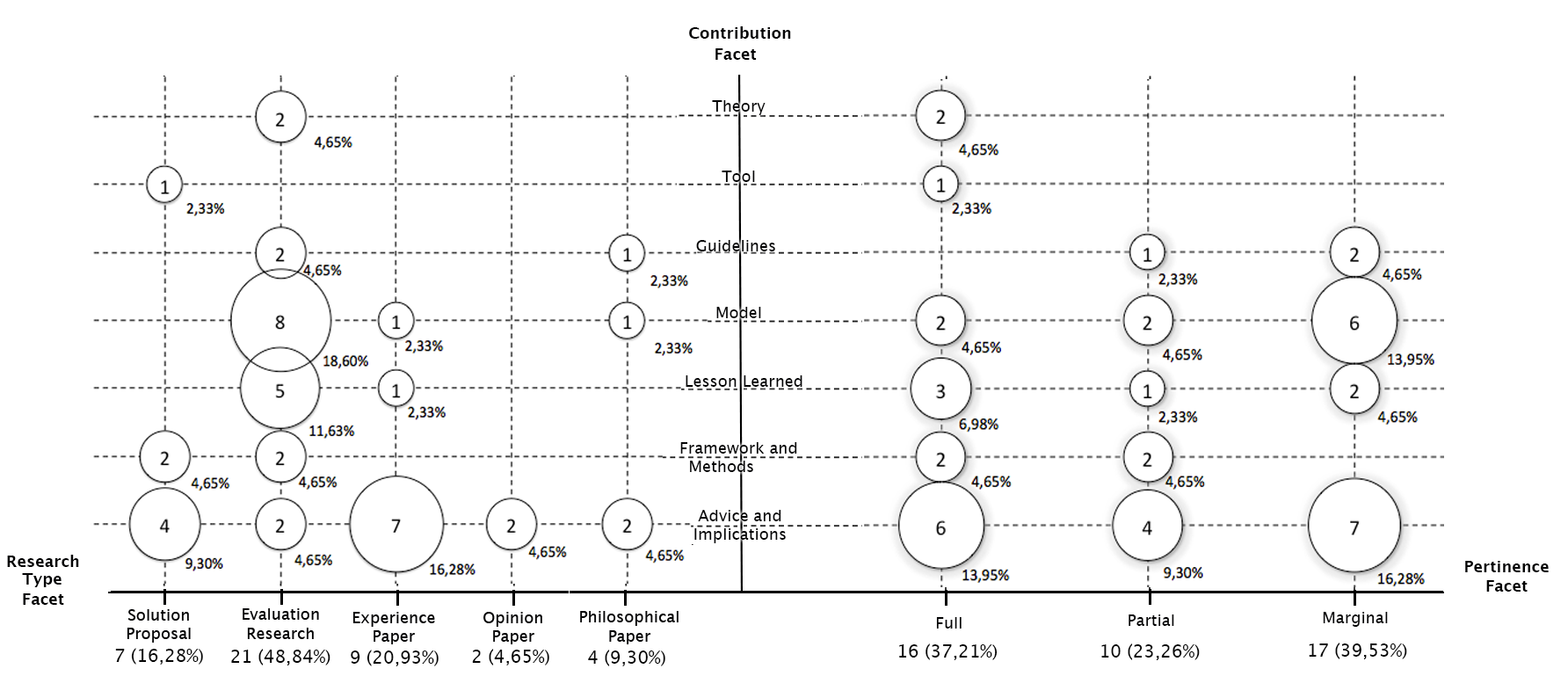}
\caption{Systematic map - \textit{Contribution}, \textit{pertinence} and \textit{research type}}
\label{fig:ms:cpr}
\end{figure*}

\begin{figure*}
\centering
\includegraphics[width=0.9\textwidth,keepaspectratio=true]{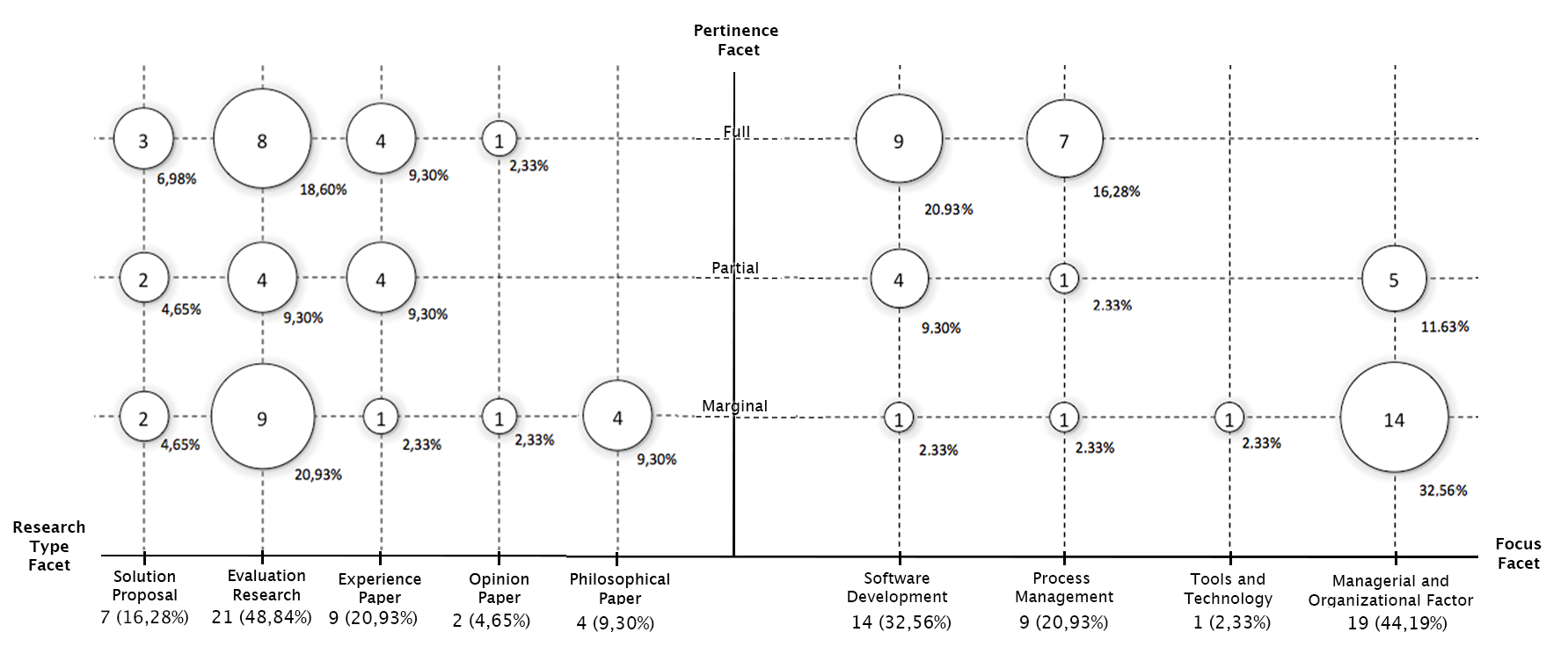}
\caption{Systematic map - \textit{Pertinence}, \textit{focus} and \textit{research type}}\label{fig:ms:pfr}
\end{figure*}

In order to illustrate potential gaps in startup research, we present the 
systematic map with multi-dimensional bubble charts (``x-y scatter plots with
bubbles in categories intersections''~\cite{Petersen2007}), where the size of 
the bubble is determined by the number of publications corresponding to the 
x-y coordinates. Differently from other studies 
(e.g.~\cite{Smite2009,Dyba2006}), each data point is represented by four
features. Thus, we created three plots 
(Figures~\ref{fig:ms:fcr}~-~\ref{fig:ms:pfr}) to visualize all six possible
facets combinations from the classification schema, giving a complete overview 
of the systematic map and providing means to analyze it. 

For example, Figure~\ref{fig:ms:fcr} indicates that 11 studies (26\% of the 
total) are focused on managerial and organizational factors, conducted through 
an evaluation type research. In the same figure it is possible to observe that 
8 studies with managerial and organizational focus contributed to the body of 
knowledge with a model. However, by looking at Figure~\ref{fig:ms:cpr}, one can 
quickly notice that 6 out of the total 10 models have only a marginal 
pertinence with engineering activities in software startups.

%MUN2: General fixes to figures:
% - numbers in bubbles need bigger font size / 
% percentage numbers should be at least current size of frequency numbers %NP TODO
% - remove frequency/percentage from x-axis (i.e. the "37 (100%)", the numbers  
%   for each item in the facet (e.g. 6 (16.22%)) can stay %NP DONE OK

%Figures need the following additional fixes (see below for each figure)
%%startFigure
%%end figure
% - x-axis rename: Research category facet -> Research type facet %NP DONE OK
% - x-axis rename: Contribution type facet -> Contribution facet  %NP DONE OK
% - x-axis rename: hilosophical -> Philosophical %NP DONE OK

 %%startFigure
%%end figure
% - x-axis rename: Research category facet -> Research type facet %NP DONE OK
% - y-axis rename: Contribution type facet -> Contribution facet %NP DONE OK

 %%startFigure
%%end figure

% subsection publication_distribution (end)

\subsection{Context characteristics of startups} % (fold)
\label{sub:contextual_features_of_startups}

To illustrate how authors use the term ``software startup'', we systematically 
extracted themes which characterize the companies in the selected primary
studies. We were able to identify 15 main themes, reported in 
Table~\ref{tab:ms:themes}. 

\begin{table*}[t]
\caption{Mapping Study - Recurrent themes }\label{tab:ms:themes}
\centering
\footnotesize
\begin{tabular}{p{0.2in}p{1in}p{3.2in}p{0.47in}p{1.5in}}
\toprule
ID & Theme & Description & Frequency & Ref. \\
\midrule
\midrule
T1 & Lack of resources & Economical, human, and physical resources are 
extremely limited. & 18 &
\cite{Yogendra2002,Yoffie1999,Crowne2002,Kajko-Mattsson2008,
Camel1994a,Coleman2008a,Coleman2007,Coleman2008,Hanna2010,Sutton2000,Ambler2002,
Stanfill2007,Tanabian2005,20124415623309,6298105,bosch2013early,
ISI:000317949700032,ISI:000312482500005} \\
T2 & Highly Reactive & Startups are able to quickly react to changes of the
underlying  market, technologies, and product (compared to more established 
companies) & 17 & 
\cite{Zettel2001,Kajko-Mattsson2008,Camel1994a,Coleman2008a,Coleman2008,
Tingling2007,Sutton2000,Fayad1997,Ambler2002,Kuvinka2011,Deias,Silva2005,
20124415623309,ISI:000312482500005,ISI:000317949700032,bosch2013early,6298105} 
\\
T3 & Innovation & Given the highly competitive ecosystem, startups need to 
focus on highly innovative segments of the market. & 15 & 
\cite{Heitlager2007,Yogendra2002,Mirel2000,Steenhuis2008,Jansen2008,Lai2010,
Sutton2000,Hasel2010,Midler2008,Kakati2003,6298105,bosch2013early,
ISI:000317949700032,ISI:000312482500005,20124315605867} \\
T4 & Uncertainty & Startups deal with a highly uncertain ecosystem under 
different perspectives: market, product features, competition, people and 
finance. & 14 & 
\cite{Heitlager2007,Kim2005,Coleman2008a,Coleman2007,Coleman2008,Fayad1997,
Midler2008,Tanabian2005,Hilmola2003,20124315605867,ISI:000312482500005,
ISI:000317949700032,bosch2013early,6298105} \\
T5 & Rapidly Evolving & Successful startups aim to grow and scale rapidly. & 
14 & 
\cite{Yogendra2002,Yoffie1999,Camel1994a,Coleman2008,Su-Chan2007,Sutton2000,
Ambler2002,Kuvinka2011,Deakins2005,20124415623309,ISI:000312482500005,
ISI:000317949700032,bosch2013early,6298105}\\
T6 & Time-pressure & The environment often forces startups to  release fast
and to work under constant pressure (terms sheets, demo days, investors' 
requests) & 13 & 
\cite{Zettel2001,Camel1994a,Coleman2008,Tingling2007,Sutton2000,Deakins2005,
Hilmola2003,Mater2000,20124415623309,ISI:000312482500005,ISI:000317949700032,
bosch2013early,6298105} \\
T7 & Third party dependency & Due to lack of resources, to build their product, 
startups heavily rely on external solutions: External APIs, Open Source 
Software, outsourcing, COTS, etc. & 10 & 
\cite{Yoffie1999,Wood2005,Wall2001,Jansen2008,Lai2010,Hanna2010,Sutton2000,
Ambler2002,6298105,bosch2013early} \\
T8 & Small Team & Startups  start with a small numbers of individuals. & 9 &
\cite{Zettel2001,Yoffie1999,Crowne2002,Kajko-Mattsson2008,Sutton2000,Chorev2006,
Tanabian2005,6298105,20124415623309} \\
T9 & One product & Company's activities gravitate around one product/service
only. & 9 & 
\cite{Ambler2002,Kuvinka2011,Bean2005,Deias,Silva2005,Coleman2008,Taipale2010,
6298105,20124415623309} 
\\
T10 & Low-experienced team & A good part of the development team is formed by 
people with less than 5 years of experience and often recently graduated 
students. & 8 & 
\cite{Camel1994a,Coleman2008a,Coleman2007,Ambler2002,Kakati2003,6298105,
bosch2013early,ISI:000312482500005} \\
T11 & New company & The company has been recently created. & 7 & 
\cite{Crowne2002,Yoffie1999,Camel1994a,Chorev2006,20124415623309,
ISI:000317949700032,bosch2013early} \\
T12 & Flat organization & Startups are usually founders-centric and everyone 
in the company has big responsibilities, with no need of high-management. & 5 & 
\cite{Yoffie1999,Kajko-Mattsson2008,Silva2005,Tanabian2005,6298105} \\
T13 & Higly Risky & The failure rate of startups is extremely high. & 5 & 
\cite{Heitlager2007,Kajko-Mattsson2008,Camel1994a,Tanabian2005,
ISI:000317949700032} \\
T14 & Not self-sustained & Especially in the early stage, startups need 
external funding to sustain their activities (Venture Capitalist, Angel 
Investments, Personal Funds, etc.). & 3 & 
\cite{Zettel2001,20124315605867,Hanna2010} \\
T15 & Little working history & The basis of an organizational culture is 
not present initially. & 3 & \cite{Yoffie1999,Ambler2002,6298105} \\
\bottomrule
\end{tabular}
\end{table*}

When discussing software startups, 18 authors reported a general lack of
human, physical and economical resources (T1). For this reason, startups
deeply depend upon external software solutions such as third party APIs, COTS
and OSS (T7). Other studies refer to  companies which are able to quickly
react to changes in the market and technologies (T2), under remarkably
uncertain conditions (T4). Some authors indicate that these companies are
focused on highly innovative segments of the market (T3), generally working
on a single core-product (T9) under extremely high time-pressure (T6).
Furthermore, 14 authors write about startups as fast growing companies
(T5) designed to rapidly scale-up. Other researches mention a very small
founding team (T8), which is often composed by low-experienced people (T10)
with a very flat organization structure (T12), where the CEO is sometimes a
core developer himself. Finally, other studies agree on the highly risky nature
of these businesses (T13), being newly created (T11) and therefore with no or 
little working history (T15).

% subsection contextual_features_of_startups (end)

\subsection{Rigor and relevance} % (fold)
\label{sub:rigor_and_relevance_results}

Even though the scientific value of a study is not determined by the 
publication type, the peer-review process required for publishing a journal 
article is generally much more rigorous and formal than the procedure to get an 
article published in a scientific magazine or accepted to a 
conference~\cite{ColinRobson2009}. Twenty (47\%) of the selected 43 primary 
studies are journals, while 16 (37\%) are published in conference proceedings 
and 7 (16\%) in magazines. Although this feature alone is not enough to 
represent a direct implication on the quality\footnote{The publication criteria 
are determined by the specific editor of the journal/magazine or the committee 
of a conference. There is a vast multitude of excellent quality studies 
presented in conference proceedings and magazines, and many examples of 
poor-quality journal articles.}, it can be interpreted as a first indicator of 
scientific quality. We formally assessed the quality of the primary studies with 
the rigor and relevance process described in 
Subsection~\ref{sub:rigor_and_relevance}, resulting in 
Figure~\ref{fig:ms:rigor-rel-map} (the raw data for this figure is available 
in the supplementary material~\cite{paternoster_supplementary_2013}).
 
Looking at Figure~\ref{fig:ms:rigor-rel-map}, nine studies (21\%) lie in the 
upper right quadrant, the preferable region, of the chart (rigor $\geq 2$, 
relevance $\geq 3$). Twenty-one studies (49\%) exhibit moderate industry 
relevance (relevance $\geq 2$), showing however low scientific rigor (rigor $\leq 
1.5$). Ten studies (23\%) are located in the lower left quadrant of the chart 
(rigor $\leq 1.5$ and relevance $\leq 2$).

%%startFigure
\begin{figure}[t]
\centering
\includegraphics[width=0.50\textwidth,keepaspectratio=true]{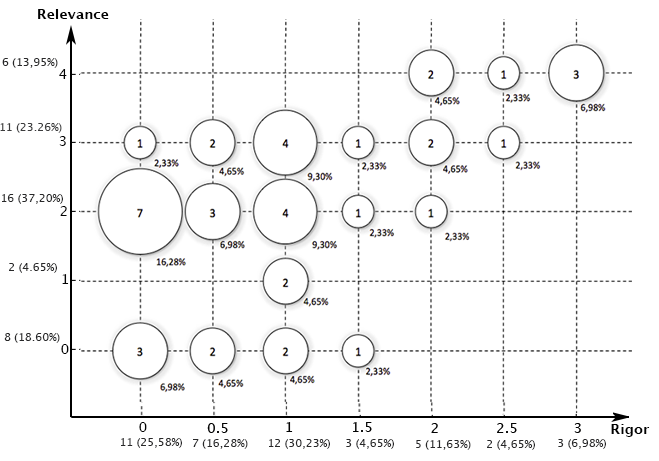}
\caption{Rigor-relevance overview}\label{fig:ms:rigor-rel-map}
\end{figure}
%%end figure

% subsection rigor_and_relevance (end)

\section{Analysis of the state-of-art} %-------- ## ----------------- ## ----------------- ## ----------------- ## ----------------- ## ---------
 \label{sect:discussion}

More than 65\% of the 43 identified primary studies have been published in the 
last ten years (between 2004 and 2013, see Figure~\ref{fig:ms:year}). Fourteen 
primary studies, dated prior 2004, discuss software startup related issues. 
This indicates that the research on startups is still in its infancy, compared 
to the long-standing history of the SE discipline, and gaining interest in the 
research community. 
The yearly distribution of publications attest the novelty of the startup
phenomenon, enabled and amplified by the potentially large markets and 
distribution channels offered by internet and mobile
devices~\cite{Christensen1997,Kakati2003}. This has opened a set of new 
challenges that are difficult to address using traditional 
approaches~\cite{Sutton2000}.

%MUN3 Deleted since does not add much
% To make a comparison, when looking at one of
% the closest domain such as software engineering in small companies, we can
% observe how the specific literature started many years before 2002. For
% example, in 1994 Brodman was already discussing how to adapt the CMM
% methodology to small organizations \cite{Brodman1994}. Only 10 years later we
% find an empirical study inquiring startups and development methodologies
% \cite{Silva2005}.

By analyzing the three bubble charts 
(Figure~\ref{fig:ms:fcr}~-~\ref{fig:ms:pfr}), the following observations can be 
made on the state-of-art:

\begin{enumerate}[\textbullet]
\item Looking at the pertinence facet in Figure~\ref{fig:ms:pfr} we can observe 
that only 16 studies (37\%) are entirely dedicated to software development in 
startups; Ten of those produced a weak contribution (advice and implications 
(6); lesson learned (3); tool (1)).
\item  Observing the focus facet (Figure~\ref{fig:ms:pfr}), it is easy to see 
that 19 studies (44\%) are focused on managerial and organizational factors. 
None of those 19 studies exhibits a full pertinence to engineering activities 
in startups.
\item The overall studies' contribution types are for the greater part weak: 
advice and implication, lessons learned, tools and guidelines (27 studies, 
63\%, Figure~\ref{fig:ms:fcr}). Of the 16 remaining studies (37\%) which 
exhibit a strong contribution type (theory, framework/method, model), only 7 
focus on what we considered fundamental for our research questions (software 
development and process management).
\item Approximately half of the selected studies were carried out using an 
evaluation type research (21 studies, 49\%, Figure~\ref{fig:ms:fcr}), being 
the only research type which involves a field study. However, we can also 
observe in the same figure that 11 of these are related to managerial and 
organizational factors, and only 8 out of 21 have a full pertinence with 
engineering activities in software startups (Figure~\ref{fig:ms:pfr}).
\item Looking at the studies which focus on process management and software 
development (Figure~\ref{fig:ms:pfr}), the majority (16 out of 23 studies) has 
a full pertinence with engineering activities in software startups.
\end{enumerate}

To summarize the systematic map, we can state that Coleman and 
O'Connor~\cite{Coleman2007,Coleman2008,Coleman2008a}, and Kajko-Mattson and 
Nikitina~\cite{Kajko-Mattsson2008} represent the strongest contributions to 
the field of startup research, considering strength of contribution type, 
pertinence to engineering activities and strength of empirical evidence. 
However, it must be noted that the three publications by Coleman and 
O'Connor are based on the same dataset originating from 21 companies.

\subsection{RQ1 - The context characterizing software development in startups} % (fold)
\label{sub:defining_startups}
The results (Table~\ref{tab:ms:themes}) indicate that there is no agreement 
on a standard definition, specifying the characteristics of a ``startup''. 
Different authors provide varying definitions and use the term ``startup'' 
referring to varying contexts. This renders any attempt to identify a solid and 
coherent body of knowledge on startup research very challenging. Looking at 
Suttons startup characterization~\cite{Sutton2000} from 2000, we can observe 
that our understanding of the nature of startups expanded to aspects such as 
innovation, fast growth, time pressure, third party dependency, focus on one 
product and flat organizational structures (see Table~\ref{tab:ms:themes}).

Defining what makes a software startup unique is an interesting problem.
Apparently, the definition is not strictly related to the size of the company. 
For instance, some authors call ``startups'' companies with 6 
employees~\cite{Mirel2000}, whilst others refer to ``startups'' with more than 
300 employees~\cite{Yoffie1999,Coleman2008}. Uniqueness is not defined through 
the age of the company alone: some authors studied startups which have been
operating for many years~\cite{Lai2010}, while others are more strict and limit 
the definition to only recently founded companies~\cite{Kajko-Mattsson2008}. 
Other authors treat ``start-up'' as a stage of a 
company~\cite{Crowne2002,Tanabian2005}.
Others claim that startups work on innovative products, without providing an
exact definition of ``innovation'', rendering this characterization less 
useful (a recent systematic study identified  ``41 definitions of innovation in 
204 selected primary SE studies''~\cite{Ali2010a}).

The most frequent reported themes concern the general lack of resources, high
reactiveness and flexibility, innovation, uncertain conditions, time pressure 
and fast growth. Since the contextual boundaries of startups resulted to be 
highly blurred, it is the researchers' responsibility who refer to ``startups'' 
to explicitly define the features of the company under study (e.g. company age, 
team size, product type, product development time).

\begin{table}[t]
\caption{Ranking of primary studies considering 
\textbf{P}ertinence, \textbf{Ri}gor and \textbf{Re}levance, 
Publication \textbf{A}ge and \textbf{T}ype, \textbf{C}ontribution type, 
\textbf{R}esearch type, and \textbf{F}ocus} 
\label{tab:ms:rig-rel-rank}
\centering
\footnotesize
\begin{tabular}{p{0.96in}p{0.15in}p{0.08in}p{0.08in}p{0.08in}p{0.08in}p{0.08in}p
{0.08in}p{0.08in}p{0.08in}}
\toprule
1st Author (year) & \textbf{Score} & P & Ri & Re & A & T & C & R & F \\
\midrule
\midrule
Coleman (2007)~\cite{Coleman2007}& \textbf{9.70} & 2.50 & 1.75 & 1.75 & 1.20 & 
1.00 & 0.50 & 0.50 & 0.50 \\
Coleman (2008)~\cite{Coleman2008a} & \textbf{9.70} & 2.50 & 1.75 & 1.75 & 1.20 
& 1.00 & 0.50 & 0.50 & 0.50 \\
Coleman (2008)~\cite{Coleman2008} & \textbf{9.70} & 2.50 & 1.75 & 1.75 & 1.20 
& 1.00 & 0.50 & 0.50 & 0.50\\
Kajko (2008)~\cite{Kajko-Mattsson2008} & \textbf{8.09} & 2.50 & 0.88 & 1.31 & 
1.20 & 0.70 & 0.50 & 0.50 & 0.50 \\
H\"{a}sel (2010)~\cite{Hasel2010} & \textbf{7.47} & 0.75 & 1.17 & 1.75 & 1.50 & 
1.00 & 0.50 & 0.50 & 0.30 \\
Hanna (2010)~\cite{Hanna2010} & \textbf{7.47} & 0.75 & 1.17 & 1.75 & 1.50 & 
1.00 & 0.50 & 0.50 & 0.30 \\
Deakins(2005)~\cite{Deakins2005} & \textbf{6.87} & 1.25 & 1.46 & 1.31 & 0.90 & 
1.00 & 0.50 & 0.15 & 0.30 \\
Camel (1994)~\cite{Camel1994a} & \textbf{6.61} & 1.25 & 1.46 & 1.75 & 0.15 & 
0.70 & 0.30 & 0.50 & 0.50 \\
Silva (2005)~\cite{Silva2005} & \textbf{6.58} & 2.50 & 0.00 & 0.88 & 0.90 & 
1.00 & 0.30 & 0.50 & 0.50 \\
Midler (2008)~\cite{Midler2008} & \textbf{6.55} & 1.25 & 0.58 & 1.31 & 1.20 & 
1.00 & 0.40 & 0.50 & 0.30 \\
Taipale (2010)~\cite{Taipale2010} & \textbf{6.53} & 2.50 & 0.00 & 0.88 & 1.50 & 
0.70 & 0.30 & 0.15 & 0.50 \\
Chorev (2006)~\cite{Chorev2006} & \textbf{6.43} & 0.75 & 1.17 & 1.31 & 0.90 & 
1.00 & 0.50 & 0.50 & 0.30 \\
Zettel (2001)~\cite{Zettel2001} & \textbf{6.32} & 2.50 & 0.58 & 0.44 & 0.60 & 
1.00 & 0.40 & 0.30 & 0.50 \\
Jansen (2008)~\cite{Jansen2008} & \textbf{6.25} & 1.25 & 0.58 & 1.31 & 1.20 & 
0.60 & 0.30 & 0.50 & 0.50 \\
Sutton (2000)~\cite{Sutton2000} & \textbf{6.11} & 2.50 & 0.58 & 0.88 & 0.60 & 
0.60 & 0.30 & 0.15 & 0.50 \\
Heitlager (2007)~\cite{Heitlager2007} & \textbf{6.08} & 2.50 & 0.58 & 0.00 & 
1.20 & 0.70 & 0.30 & 0.30 & 0.50 \\
Tingling (2007)~\cite{Tingling2007} & \textbf{5.99} & 2.50 & 0.29 & 0.00 & 1.20 
& 0.70 & 0.30 & 0.50 & 0.50 \\
Deias  (2002)~\cite{Deias} & \textbf{5.92} & 2.50 & 0.29 & 0.88 & 0.60 & 0.70 & 
0.30 & 0.15 & 0.50 \\
Stanfill (2007)~\cite{Stanfill2007} & \textbf{5.74} & 0.75 & 0.88 & 1.31 & 1.20 
& 0.70 & 0.30 & 0.30 & 0.30 \\
Wood (2005)~\cite{Wood2005} & \textbf{5.70} & 1.25 & 0.29 & 1.31 & 0.90 & 1.00 
& 0.30 & 0.15 & 0.50 \\
Steenhuis (2008)~\cite{Steenhuis2008} & \textbf{5.65} & 0.75 & 0.58 & 1.31 & 
1.20 & 0.70 & 0.30 & 0.50 & 0.30 \\
Yogendra (2002)~\cite{Yogendra2002} & \textbf{5.55} & 1.25 & 0.58 & 1.31 & 0.60 
& 0.70 & 0.30 & 0.50 & 0.30 \\
Ambler (2002)~\cite{Ambler2002} & \textbf{5.53} & 2.50 & 0.00 & 0.88 & 0.60 & 
0.60 & 0.30 & 0.15 & 0.50 \\
Crowne (2002)~\cite{Crowne2002} & \textbf{5.48} & 2.50 & 0.58 & 0.00 & 0.60 & 
0.70 & 0.30 & 0.30 & 0.50 \\
Mater (2000)~\cite{Mater2000} & \textbf{5.45} & 1.25 & 0.29 & 1.31 & 0.60 & 
0.70 & 0.50 & 0.50 & 0.30 \\
Kakati (2003)~\cite{Kakati2003} & \textbf{5.41} & 0.75 & 0.58 & 0.88 & 0.90 & 
1.00 & 0.50 & 0.50 & 0.30 \\
Kuvinka (2011)~\cite{Kuvinka2011} & \textbf{5.28} & 1.25 & 0.00 & 0.88 & 1.50 & 
0.70 & 0.30 & 0.15 & 0.50 \\
Li (2007)~\cite{Su-Chan2007} & \textbf{5.26} & 0.75 & 0.00 & 1.31 & 1.20 
& 0.70 & 0.30 & 0.50 & 0.50 \\
Lai (2010)~\cite{Lai2010} & \textbf{5.23} & 0.75 & 0.00 & 0.88 & 1.50 & 
1.00 & 0.30 & 0.50 & 0.30 \\
Mirel (2000)~\cite{Mirel2000} & \textbf{4.98} & 1.25 & 0.35 & 0.88 & 0.60 & 
1.00 & 0.30 & 0.30 & 0.30 \\
Himola (2003)~\cite{Hilmola2003} & \textbf{4.57} & 0.75 & 0.58 & 0.44 & 0.90 & 
1.00 & 0.30 & 0.30 & 0.30 \\
Kim (2005)~\cite{Kim2005} & \textbf{4.53} & 0.75 & 0.88 & 0.00 & 0.90 & 0.70 & 
0.50 & 0.50 & 0.30 \\
Wall (2001)~\cite{Wall2001} & \textbf{4.28} & 1.25 & 0.00 & 0.88 & 0.60 & 0.60 
& 0.30 & 0.15 & 0.50 \\
Yoffie (1999)~\cite{Yoffie1999} & \textbf{4.22} & 0.75 & 0.29 & 0.88 & 0.60 & 
0.60 & 0.30 & 0.50 & 0.30 \\
Bean (2005)~\cite{Bean2005} & \textbf{3.50} & 0.75 & 0.00 & 0.00 & 0.90 & 1.00 
& 0.30 & 0.15 & 0.40 \\
Tanabian (2005)~\cite{Tanabian2005} & \textbf{3.10} & 0.75 & 0.00 & 0.00 & 0.90 
& 0.70 & 0.30 & 0.15 & 0.30 \\
Fayad (1997)~\cite{Fayad1997} & \textbf{2.45} & 0.75 & 0.00 & 0.00 & 0.15 & 
0.60 & 0.30 & 0.15 & 0.50 \\
\bottomrule
\end{tabular}
\end{table}

\subsection{RQ2 - Transferability of results to 
industry}\label{sub:transferability}

Figure~\ref{fig:ms:rigor-rel-map} illustrates a major weakness of the 
state-of-art in startup research. Seven primary studies (16\% of the 
total), received an average score for industrial relevance ($2$) but a low 
score ($0$) for scientific rigor. According to the authors of the 
rigor-relevance model~\cite{Ivarsson2010}, this represents a major threat to 
the transferability of the results to industry. Even though findings may
appear to be somewhat appealing to practitioners (average relevance), low 
scientific rigor will render knowledge transfer to industry highly unlikely or 
even dangerous. One of the most important factors contributing to academic 
results being applied in the industry is the provision of strong scientific 
evidence~\cite{Kitchenham2005,Kitchenham2004}.

%MUN3 The first part of the analysis below I don't understand, the second part 
%about author background stands on weak ground... deleted for now but can be 
%added if refined
% Finally, by looking at how bubbles are distributed on the chart (see Figure
% \ref{fig:ms:rigor-rel-map}) with no need of sophisticated statistical methods,
% it is clear the relation between items which received the highest scores for
% \textit{rigor} and the ones which received the highest score for
% \textit{relevance}. Since the two dimensions represent two independent
% variables the result is not straightforward. In this particular sample, it
% indicates that results of studies, which have been rigorously reported, are
% generally more relevant for the industry (and vice-versa). By thoroughly
% analyzing the studies in the high rigor-relevance region, we observed that the
% authors have strong academic backgrounds but at the same time they have worked
% for many years in direct contact with startups. This is even more visible when
% comparing them to the other authors in the sample with lower value of rigor
% and relevance\footnote{Since it is not possible to obtain a complete
% background study of each author, we based our assumption on a small biographic
% review.}. This could somehow explain the reason why, in our sample, the
% rigorous studies are also highly relevant.

In the remainder of this subsection we extend the analysis of rigor and 
relevance by integrating factors such as publication type and year, but also 
the classification schema. We follow the procedure described in 
Subsection~\ref{sub:rigor_and_relevance}, computing a score (in the range
$[0-10]$) for each primary study. 
% Table~\ref{tab:ms:score-weights} lists the
% dimensions which were used to assign the final score, next to the weight to 
% balance the ranking criteria.

The design of the ranking function considers our research question of 
identifying software engineering work practices in startups. Hence, the 
pertinence dimension from the classification schema contributes the most (25\%) 
to the score, followed by rigor and relevance (17.5\% each). Age of the 
publication (15\%) is factored in as more recent studies are likely to provide 
a more relevant context for practitioners. The publication type accounts for 
10\%. The contribution type, research type and focus account each for 5\% of the total 
score which is the sum of all eight weighed scores. The conversion 
tables to achieve a normalized score are available in the supplementary 
material~\cite{paternoster_supplementary_2013}. 

Table~\ref{tab:ms:rig-rel-rank} quantifies the body of knowledge on startup 
research, provided by 43 primary studies analyzed in this systematic mapping 
study. The ranking gives an indication to what extent we can answer questions 
targeted at the state-of-art of the software engineering work practices in 
startups.

\section{RQ3 - Work Practices in startups}
\label{sect:work_practices}

We have extracted a total of 213 work practices\footnote{Note that this 
number does not reflect unique work practices but the total number; a 
detailed table of work practices is available in the supplementary 
material~\cite{paternoster_supplementary_2013}.}
from the 43 primary studies reviewed in this SMS and subsequently divided them
in categories (Table~\ref{tab:wp:extracted}), as explained in
Subsection~\ref{sub:synthesis}. The categorization of working practices 
is defined according to the focus facet of the classification schema, presented 
in Figure~\ref{subtab:focus}.
In the remainder of this section, we discuss the identified work practices, 
pointing out where gaps exist and further research is warranted.

\begin{table}[H]
\caption{Categorization of the identified work 
practices}\label{tab:wp:extracted} 
\footnotesize  
\centering 
\begin{tabular}{ll}
\toprule Software Development (Subsection~\ref{sub:swdevprac}) & 90 \\
Managerial/organizational (Subsection~\ref{sub:analysis-managerial}) & 70 \\
Process management (Subsection~\ref{sub:analysis-process}) & 47 \\ Tools and
technologies (Subsection~\ref{sub:analysis-other}) & 6 \\ \midrule Sum & 213\\
\bottomrule 
\end{tabular} 
\end{table}

% subsection software_development_in_startups_review (end)

\subsection{Process management practices}
\label{sub:analysis-process}

Process management represents all the engineering activities used to
manage product development in startups. Sutton~\cite{Sutton2000}
recognized the need for flexibility to accommodate frequent changes in the
development environment, and for reactiveness to obtain timely response
in applying methodologies. 

Agile methodologies have been considered the most viable processes for software 
startups, given that Agile methodologies embrace changes rather than avoiding 
them, allowing development to follow the business strategy~\cite{Taipale2010}. 
In this context, fast releases with an iterative and incremental approach 
shorten the lead time from idea conception to production with fast 
deployment~\cite{Taipale2010}. The benefits of having weekly releases and 
frequent build cycles, addressing the uncertainty of the market, has been 
further reported by Blank~\cite{ISI:000317949700032}, 
Tingling~\cite{Tingling2007}, Ambler~\cite{Ambler2002} and
Silva~\cite{Silva2005}.

A variant to Agile has been the Lean 
Startup~\cite{Ries2011,ISI:000317949700032}, which advocates the identification 
of the most risky parts of a software business and provide a minimum viable product (MVP) to 
systematically test and plan modification for a next iteration. In this regard, 
in order to shorten time-to-market, prototyping is
essential~\cite{Deakins2005,Camel1994a}. To allow better prototyping activities,
evolutionary workflows are needed to implement "soft-coded" solutions in the
first phases until the optimal solution is found~\cite{Deakins2005,Sutton2000}.

Coleman~\cite{Coleman2007} reports that XP is the most used development
methodology across startup companies, due to its reduced process costs and low
documentation requirements. Also other agile practices are explored, such as 
Scrumban~\cite{Kuvinka2011}, but not rigorously researched. In any case, flexible in nature, 
startups' processes don't strictly follow any specific methodology, 
but opportunistically select practices (e.g. pair-programming~\cite{Deias}, 
backlog~\cite{Tingling2007}). In fact, processes are tailored to the specific
features that characterize each development context~\cite{6298105,Coleman2007,
Coleman2008}. For example, Bosch et al.~\cite{bosch2013early} advocate for
adjusting the Lean startup methodology to accommodate the development of
multiple ideas and integrate them when time for their testing and validation
is too long. This concurs with the practice of allocating varying effort for
formalizing specifications, design, documentation and testing in tailored
development methodologies~\cite{Ambler2002,Zettel2001,Camel1994a}, emphasizing
the importance of minimal process management.

Summarizing, process management practices, reported to be useful in startups, are:

\begin{enumerate}[\textbullet]
\item Light-weight methodologies to obtain flexibility in choosing tailored
practices, and reactiveness to change the product according to business
strategies.
\item Fast releases to build a prototype in an evolutionary fashion and quickly 
learn from the users' feedback to address the uncertainty of the market.
\end{enumerate}

\subsubsection*{Discussion}
The Cynefin framework~\cite{Kurtz2003} can be used to explain the orientation
of startups towards flexible and reactive development approaches. Within this
framework, startups cross the complex and chaotic domains. Those two domains
represent the areas where applying rigorous process management to control
development activities is not effective, because no matter how much time is
spent in analysis, it is not possible to identify all the risks or accurately
predict what practices are required to develop a product.
Instead, flexible and reactive methods,
designed to stimulate customer feedback, increase the number of perspectives
and solutions available to decision makers. Moving from complex to chaotic 
domains, software startups open up new possibilities for creation, generating 
the condition for innovations. Therefore, any process tailored to the startup 
context needs at least to allow, but optimally even facilitate movements 
between complex and chaotic domains that are intrinsic in the innovation 
generation of startups. In our opinion, this is the main requirement for future 
attempts of adapting software engineering processes to the startup context.

Developers should have the freedom to choose activities quickly, stop 
immediately when results are wrong, fix the approach and learn from previous 
failures. In this regard, in line with the Lean Startup movement, we expect 
methodologies and techniques tailored from common Agile practices to 
specific startups' culture and needs, where failing is completely acceptable, 
even preferred in favor of a faster learning process. However at some point, in 
preparation for growth, startups need to plan for scalable processes. Similarly 
to SMEs~\cite{IJMR:IJMR116}, they need to find a balance between flexibility 
and repeatability in their organizations' knowledge management and processes.

\subsection{Software development practices}
\label{sub:swdevprac}

We have categorized work practices related to software development as 
illustrated in Table~\ref{tab:wp:extracted:sw}, discussing them individually.

\begin{table}[H]
\caption{Software development practices}\label{tab:wp:extracted:sw}
\footnotesize 
\centering
\begin{tabular}{ll}
\toprule
Requirements Engineering (Subsection~\ref{sub:swdevprac:re}) & 21 \\
Design and Architecture (Subsection~\ref{sub:swdevprac:da}) & 32 \\
Implementation, Maintenance and Deployment (Subsection~\ref{sub:swdevprac:imd}) 
& 14 \\
Quality Assurance (Subsection~\ref{sub:swdevprac:qa}) & 23 \\
\midrule
Sum & 90 \\
\bottomrule
\end{tabular}
\end{table}

\subsubsection{Requirements Engineering practices}
\label{sub:swdevprac:re}

Establishing an engineering process for collecting, defining and managing
requirements in the startup context is challenging. RE practices are often 
reduced to some key basic activities~\cite{Zettel2001,Crowne2002}. Su-Chan et 
al.~\cite{Su-Chan2007} report on efforts in defining the value-proposition that 
the company aims to achieve at the very first stage of the project.

Initially, as startups often produce software for a growing target 
market~\cite{ISI:000317949700032,Camel1994a}, customers and final users are 
usually not well-known and requirements are therefore 
market-driven~\cite{Deakins2005} rather than customer-specific. In such
a context Mater and Subramanian~\cite{Mater2000} attest severe difficulties in
eliciting and detailing the specifications of both functional and non-functional 
requirements. Moreover, in unexplored and innovative markets, the
already poorly-defined requirements tend to change very rapidly. This makes it
hard for the development team to maintain requirements and keep them
consistent over time.

Several authors acknowledge the importance of involving the customer/user in 
the process of eliciting and prioritizing requirements according to their 
primary needs~\cite{6298105,Deakins2005,Midler2008,Crowne2002,Wall2001}. 
However, the market-driven nature of the requirements demands for alternatives.
For example, startups can use scenarios in order to be able to identify 
requirements in the form of user stories~\cite{Silva2005} and 
estimate the effort for each story~\cite{Zettel2001}. In scenarios and in 
similar product-usage simulations, an imaginary customer can be represented
by an internal member of the company~\cite{Tingling2007}. An example of a more 
strict customer-development process~\cite{Adebanjo2010} that drives the 
identification of requirements can be found in an experience report by
Taipale~\cite{Taipale2010}. 

\subsubsection*{Discussion}
Polishing requirements that address an unsolicited need is waste. 
To demonstrate problem/solution fit it is required to discover the real needs 
of your first customers, testing business speculations only by defining a 
minimal set of functional requirements.
In the future, developing a deep customer collaboration, such as the customer
development process~\cite{Blank2011a} will change the requirements
elicitation methods, moving towards testing the problem and understand if the
solution fits to real needs even before the product goes to the market.

\subsubsection{Design and Architecture practices}\label{sub:swdevprac:da}
Deias and Mugheddu~\cite{Deias} observed a general lack of written
architecture and design specifications in startups. Development in 
startups is usually guided by simple principles, trying to avoid 
architectural constraints that are difficult to overcome as the product and 
user-base grows. Tinglings~\cite{Tingling2007} results suggest that a not well 
analyzed architecture causes problems in the long run. However, a good-enough
architecture analysis should at least identify the decisions that might cause 
problems before obtaining product revenue, and which can be fixed at a later 
point in time, accounting for increased resources after revenue cash starts to 
flow~\cite{6298105}.

One common analysis on determining requirements is the planning game, where
developers can arbitrate the cost of desired features and delivered
functionalities. However business people can ``steer'' the process and impact 
adopted architectural decisions, which could present obstacles for refactoring 
activities, especially if the software complexity grows~\cite{Ambler2002}. Then 
the use of design patterns~\cite{gamma_design_1994}, e.g. the 
model-view-controller~\cite{Gamma1993}, can provide advantages to startups which 
need flexibility in refactoring the product.
Moreover, formulating initial architectural strategies with high-level 
models and code-reuse from industry standards represents a viable trade-off 
between big up-front and ad-hoc design~\cite{bosch2013early}.

Jansen et al.~\cite{Jansen2008} suggest that startups should
take advantage of existing components, reducing thereby time-to-market. 
Leveraging on frameworks and code-reuse of third party components reinforces 
the architectural structure of the product and the ability to scale with the 
company size. As reported by Yoffie~\cite{Yoffie1999}, scalability 
represents the most important architectural aspect in startups and should be 
addressed as soon as possible. Then, startups can benefit from reusing 
components and shared architectures across projects as well as existing 
high-level frameworks and modeling practices.

Summarizing, design and architectural practices reported to be useful in
startups are:
\begin{enumerate}[\textbullet]
\item The use of well-known frameworks able to provide fast changeability of 
the product in its refactoring activities.
\item The use of existing components, leveraging third party code reinforcing
the product ability to scale.
\end{enumerate}

\subsubsection*{Discussion}
%MUN This "speculation" is ok, but you need to state explicitly whether your 
%suggestions can be implemented immediately because the techniques/methods 
%exists (then you need to refer to them) OR they cannot be realized immediately 
%since they don't exist and research is needed (then you need to state that 
%EXPLICITLY too). Now you just suggest to do something, but it is not clear 
%HOW and whether it is possible NOW.
Despite the general lack of written architecture specifications in startups,
the difficulties presented when the user-base and product complexity grows
can be overtaken by a little upfront effort on framework selection and
analyzing decisions that might cause problems before obtaining product
revenue.
When the product evolves, the use of architecture and design to make features 
modular and independent are crucial to remove or change functionalities. 
Therefore, employing architectural practices and frameworks
that enable easy extension of the design (e.g. pluggable architecture where
features can be added and removed as plugins~\cite{plug}) can better align the 
product to
the uncertainty of the market needs.

\subsubsection{Implementation, Maintenance and Deployment practices} 
\label{sub:swdevprac:imd}

Silva and Kon~\cite{Silva2005} report on positive results from pairing up senior
and junior developers together in pair-programming sessions. During these
sessions, developers also made use of coding standards to help cross-team code
readability and reduce the risks related to the high-rate of developer turnover.
In a different study, Tingling~\cite{Tingling2007} attested an  initial 
high resistance to the introduction of coding standards and  pair-programming. 
These practices were then adopted only in later stages,  when the complexity of 
the project required them. Zettel et al.~\cite{Zettel2001}  report that the 
practice of tracking traditional code metrics has been labeled as  ``obsolete 
and irrelevant'' and that many companies use their ad-hoc methods of assessing
processes and metrics.   The software team studied by Zettel et
al.~\cite{Zettel2001} had a bug-fix process centered  around their 
issue-tracking tool and relied on a release system.  Several authors reported on
advantages of constant code refactoring:  ensuring readability and
understandability of the code~\cite{Zettel2001},  improving
modularity~\cite{Tingling2007} and providing discipline to
developers~\cite{Taipale2010}. On the other hand, introducing refactoring may
cause problems since developers had no or little experience~\cite{Deias} and
they didn't see immediate value in introducing high level
abstractions~\cite{Silva2005}. In the case study described by 
Ambler~\cite{Ambler2002}, an initial lack of refactoring led to the need of 
re-implementing the whole system after the number of users had grown 
drastically. Finally some authors reported work practices related to deployment 
claiming that some software teams deploy manually the code on the 
infrastructure~\cite{Silva2005} while others rely on continuous  integration 
tools~\cite{Taipale2010}.

\subsubsection*{Discussion}
Startups tend to start the code implementation with an informal
style, introducing standards only when the project size requires them. This 
is in line with the observations made by Thorpe et al.~\cite{IJMR:IJMR116} on 
knowledge management and growth in SMEs. The process is often driven by 
lightweight ad-hoc systems: bug-tracking, simple code 
metrics and pair-programming sessions.
In this regard, startups in the early stage keep the code base small and simple 
to develop only what is currently needed, maintaining focus on core
functionalities to validate with first customers. As the business goal drives
the need of effort in refactoring and implementation, more studies will be
needed to align business with execution of concrete development practices in
startup contexts (e.g. GQM Strategies~\cite{basili2013linking}).

\subsubsection{Quality Assurance practices}
\label{sub:swdevprac:qa}

Testing software is costly in terms of calendar time and often compromised in 
startups~\cite{Camel1994a,Zettel2001}. Quality assurance, in the broader sense, 
is largely absent because of the weak methodological management, discussed 
in Subsection~\ref{sub:analysis-process}. The complex task of implementing test
practices is further hindered by the lack of team 
experience~\cite{Silva2005,Mater2000}.

However, usability tests are important to achieve product/market 
fit~\cite{Mirel2000}. Ongoing customer acceptance ensures that 
features are provided with tests, which can effectively attest the fitness of 
the product to the market~\cite{Tingling2007,6298105,ISI:000317949700032}. Mater 
and Subramanian~\cite{Mater2000} suggest to use a small group of early 
adopters or their proxies as quality assurance fit team. Furthermore, users can 
be an important means to judge whether the system needs more 
tests~\cite{Zettel2001}.
Outsourcing quality assurance to external software testing experts,
handling the independent validation work if resources are not
available~\cite{Mater2000}, can also be an alternative.

Summarizing, quality assurance practices, reported to be useful in startups, are:

\begin{enumerate}[\textbullet]
\item The use of ongoing customer acceptance with the use of a focus groups of early
adopters, which targets to attest the fitness of the product to the market.
\item Outsourcing tests if necessary, to maintain the focus on the development
of core functionalities.
\end{enumerate}

\subsubsection*{Discussion}
Even though testing software is costly, acceptance tests are the only practice 
to validate the effectiveness of the product in uncertain market
conditions~\cite{ISI:000317949700032}.
Therefore, providing time-efficient means to develop, execute and maintain 
acceptance tests is the key to improve quality assurance in
startups~\cite{Mater2000}.
In our opinion, startups will start making use of different automatic 
testing strategies, when easily accessible (e.g. create a test from each fixed 
bug~\cite{6569740}).
Considering startups' frequent changing activities during the validation of 
the product on the market, UI testing remains not a simple but important task. 
However, more research is needed to develop practical UI testing approaches 
that can be commercialized~\cite{Banerjee20131679}.

\subsection{Managerial and organizational practices} 
\label{sub:analysis-managerial} 

Flexibility, more than structure, plays an important role in
startup companies~\cite{Hilmola2003,20124415623309}. Time pressure and lack of 
resources~\cite{ISI:000312482500005,20124315605867} lead to a loose 
organizational structure and often lack of traditional project
management~\cite{Camel1994a}. To accommodate flexibility of managerial and 
organizational practices, the empowerment of team members represents the main viable 
strategy to enhance performance and chances of 
success~\cite{Camel1994a,Steenhuis2008}. The team needs to be able to absorb 
and learn from trial and error quickly enough to adapt to new emergent 
practices~\cite{Midler2008,Sutton2000,Steenhuis2008}.

Empowerment allows the team to move rapidly and cutting through the
bureaucracy, approval committees and veto 
cultures~\cite{6298105,bosch2013early}. Nevertheless, key
performance indicators (e.g. customer attrition, cycle time) and continuous
deployment processes can effectively assess the consumers' demand using the 
least amount of resources possible \cite{Ries2011}. However, in building up a 
startup company, the team needs expertise to counterbalance the lack of 
resources~\cite{Chorev2006,Coleman2008a,6298105}. Working on innovative products
requires creativity, ability to adapt to new roles and to face new challenges 
everyday~\cite{Sutton2000,Hasel2010,ISI:000312482500005}, working
overtime if necessary~\cite{Camel1994a,Tanabian2005}. Previous experience in
similar business domains~\cite{Yoffie1999,20124315605867,ISI:000312482500005} 
and a working history in a team, exhibiting characteristics of an entrepreneur 
(courage, enthusiasm, commitment, leadership), also play a primary 
role~\cite{Chorev2006,Kakati2003,20124415623309} in the skill set of a startup 
employee.

Nevertheless, the absence of structure might hinder important activities, such
as sharing knowledge and team coordination, especially when the company
grows~\cite{Ambler2002}, as also observed in the context of 
SMEs~\cite{IJMR:IJMR116}. In this case co-location is essential to facilitate
informal communication and close interactions between development and business
concerns~\cite{Coleman2007,Tingling2007}. 
Effective organization and governance mechanisms need to enable and maintain 
alignment between business and technology strategies, avoiding waste of 
resources~\cite{Yogendra2002,Zettel2001}.
Moreover, Crowne suggests to plan organizational objectives in the short-medium 
term~\cite{Crowne2002}, measuring development cycle time to find areas for 
improvements~\cite{Taipale2010,Kajko-Mattsson2008}. 

Finally, despite Camel~\cite{Camel1994a} reports lack of documentation in 
startup companies, Kuvinka~\cite{Kuvinka2011} argues that startups, when
approaching the development of long user stories, can take advantage of 
documentation and sometimes UI design to facilitate their management in the 
long run, especially when interacting with third parties~\cite{Taipale2010}. 

Summarizing, managerial and organizational practices reported to be useful in 
startups are:

\begin{enumerate}[\textbullet]
\item Empowerment of team responsibilities and their ability to influence the 
final outcomes.
\item The use of key performance indicators to assess the consumers' demand.
\item Plan of short-medium term objectives, measuring development cycle time
to find areas of improvement.
\end{enumerate}

\subsubsection*{Discussion}
%MUN Hmm, this discussion is still quite vague... we know that at least for 
%young startups, empowerment is important and necessary. Can you say something 
%in relation to the knowledge management paper I've used in the related work? 
%In particular, maybe you should discuss startup growth and the importance of 
%managing knowledge, keeping it flexible but still rigid enough to be scalable?
More empowerment allows the team to move rapidly with less bureaucracy. Then,
the development team plays a key factor to enhance commitment, creativity and
ability to adapt to new roles when necessary. In addition, open communication
remains crucial for startups to handle engineering activities, understanding
the progress, code conflicts and competences. Therefore, new tools and
techniques for focusing, exploring and making observations, encouraging
comparison and seeking clarification and validation could improve effective
verbal communication and lead to less misunderstandings and confusion.
Nevertheless, in view of growth, managing transferable knowledge becomes
crucial when hiring new personnel. However, keeping it informal but still
providing enough structure for knowledge creation is challenging. In this
regard, Thorpe et al.~\cite{IJMR:IJMR116} suggest to design a flexible
``learning architecture'', sensitive to entrepreneurial characteristics and
specific context of the company, without limiting creativity.

\subsection{Tools and technologies}\label{sub:analysis-other}
Startups are often established to develop technologically innovative products, 
which in turn might require cutting-edge development tools and techniques. 
Technological changes in the IT industry swipe through new network 
technologies, increasing variety of computing devices, new programming 
languages, objects and distribution technologies~\cite{Sutton2000}.

However, from a managerial perspective, startups still prefer easy-to-implement 
tools, such as white-boards and real-time tools that are easy to use for 
handling fast-paced changing information. Sophisticated solutions, such as CASE 
tools~\cite{Kuhn_1989}, require training and have high implementation and 
maintenance costs~\cite{Bean2005,Ambler2002}.

In general, startups take advantage of those technologies that can quickly
change the product and its management~\cite{Silva2005,Crowne2002}, avoiding 
conflicts with business strategic plans. Examples include general-purpose 
infrastructures, such as configuration management, problem reporting and 
tracking systems, planning, scheduling and notification systems. Such 
technologies support the needed activities, accommodating changes when 
required~\cite{Sutton2000}. To mitigate the lack of resources, startups might 
take advantage of open source solutions when possible, which also gives access 
to a large pool of evaluators and evolving 
contributions~\cite{Wall2001,Wood2005}.

Summarizing, tools and technologies practices reported to be useful in startups 
are:

\begin{enumerate}[\textbullet]
\item The use of easy-to-implement tools to work with fast-paced changing 
information.
\item The use of open source solutions.
\end{enumerate}

\subsubsection*{Discussion}
Startups can take advantage of using new technologies and development tools
without having any legacy or being constrained by previous working experience. 
However, lack of experience can also be a disadvantage which could be 
compensated by a light-weight process to select technologies; this selection 
could be guided by domain-specific or product specific requirements. For 
example, if a startup is creating a product that is meant to work with 
a growing amount and diversity of consumer mobile devices, to create business 
advantages, the platform should be easy to extend to support new
devices~\cite{smutny2012mobile}.

\section{Conclusions and future work} %-------- ## ----------------- ## ----------------- ## ----------------- ## ----------------- ## ---------
 \label{sect:concl}

Startups are able to produce software products with a strong impact on the
market, significantly contributing to the global economy. Software development,
being at the core of startups' daily activities, is however not supported by a
scientific body of knowledge. The evidence provided by the 43 primary studies
is, for the most part, inadequate to understand the underlying phenomenon of
software development in startups. To the current date, fourteen years after 
Sutton assessed that startups have been neglected from process studies
\cite{Sutton2000}, the gap has been only partially filled.

This is remarkable, considering startups' proliferation and, at the same time, 
high failure rate. Hence, to be able to intervene on the software development 
strategy with scientific and engineering approaches, it is necessary to better 
understand and characterize the state-of-art in the software startup context.

% What have we done 
By means of a systematic mapping study, we provide a classification of the 
state-of-art, assess rigor and relevance of the identified primary studies, and
analyze software development work practices discussed in the surveyed 
literature.

%What we found
Looking at the 43 primary studies, 19 (44\%) are focused on 
managerial and organizational factors. Only 16 studies (37\%) are entirely 
dedicated to software development in startups, whereby 10 studies constitute 
a weak contribution type. Overall, only 4 contributions 
to the field are entirely dedicated to engineering activities in startups, 
providing a strong contribution type and conducted through an evidence-based 
research 
approach~\cite{Coleman2008,Coleman2008a,Coleman2007,Kajko-Mattsson2008}. 
However, three of these studies are based on the same data, leading to the 
conclusion that there is a lack of relevant primary studies on software 
development in the startup context. In the following subsections we provide 
answers to our initially posed research questions.

\subsection{RQ1 What is the context that characterizes software development in startups?}

There is no unique definition in literature on what constitutes a startup. The 
inconsistent use of the term ``startup'' by different authors and lacking 
descriptions of context restrains the creation of a coherent body of knowledge 
on software startups. This also hinders the adoption of results by practitioners
as the context in which advancements are applicable is lacking.

The most frequently reported contextual features of startups concern the 
general lack of resources, high reactiveness and flexibility, intense 
time-pressure, uncertain conditions and fast growth. Since the contextual 
boundaries of startups resulted to be highly blurred, we believe
it is responsibility of the researchers who refer to the term ``startup'' to 
explicitly define the features of the studied companies. In most of primary 
studies an explicit contextualization has been neglected, affecting the 
generalizability of their results.

Some of the features characterizing startups are common to other SE contexts:  
innovative products development, market-driven development, small companies, 
short time-to-market, web-development. However, the coexistence of all 
these features poses a new, unique series of challenges.

\subsection{RQ2 To what extent does the research on startups provide reliable 
and transferable results to industry?}
The rigor and relevance analysis indicates that only a minority of the studies 
(9, 21\%) representing the state-of-art provide transferable and reliable 
results to practitioners. Even more concerning is that more than half of the 
studies (23, 53\%) exhibit moderate industry relevance, however at the same 
time low scientific rigor. This makes the transfer of results to practitioners 
highly unlikely or even dangerous~\cite{Ivarsson2010}, calling for more 
rigorous studies in the context of software startups.

\subsection{RQ3 What are the reported work practices in association with software
engineering in startups?}

We extracted and discussed work practices commonly used in startups as 
reported in the reviewed literature. In terms of process management, 
agile and more traditional methodologies struggle to get adopted by startups due to 
an excessive amount of uncertainty and high time-pressure. Light-weight 
methodologies that allow companies to pick and tailor practices are preferred 
as they facilitate reactiveness and allow rapid changes in the product. In this 
sense, processes in startups are evolutionary in nature, and the product is 
obtained by iterating and updating an early prototype driven by customer
feedback.

Software development practices are reported to
be adopted only partially and mostly in a late stage of the startup life-cycle.
Requirements are market-driven and hardly documented. The architecture and
design is often replaced by the use of well-known frameworks that facilitate
maintenance and reduce documentation efforts. Ad-hoc code metrics, pair 
programming sessions, and code refactoring sessions are reported to be valuable 
practices. Testing is mostly conducted through customer acceptance, focus 
groups, and sometimes by outsourcing the testing activity.
 
Managerial and organizational practices are reduced to the essential. With 
minimal bureaucracy, developers are largely empowered and encouraged to adapt 
to several roles, creatively impacting on the final products. Given the unpredictability in 
the startup context, milestones and objectives are focused on the short-medium 
term and basic key performance indicators are used to track customers' demands.

Startups mainly make use of simple tools to support and trace the 
knowledge-base and manage the workflow, often opting for open-source solutions 
that require little or no training and maintenance.

\subsection{Implications for practitioners, research and future work}
Evidence from the reviewed primary studies indicates that startup 
companies, struggling to survive and operate in an unpredictable context, can
benefit from adopting certain software engineering practices. However, low 
rigor studies and insufficient context information threaten the adoption of 
these practices by practitioners.

Performing research on startups is challenging due to the rapidly-changing 
conditions surrounding the studied environment. Therefore it is crucial to 
explicitly define the context when studying startups, describing however also 
study design and validity threats. This strengthens studies, lifting the 
potential of results transfer to industry.

While the characterization of startups through recurrent themes presented 
in this paper can serve as a basis, future work is needed to compile a common 
startup terminology and a set of definitions. That would support the generation 
of a consistent body of knowledge based on evidence-based research, aiming at 
supporting activities and decisions of the growing number of startup companies.
We are currently investigating early-stage startups that are recently founded 
and distributed in different geographic areas and market sectors. Following a 
grounded theory approach, we aim at exploring the state-of-practice in this 
context, identifying software development strategies engineered by 
practitioners.

\bibliographystyle{elsarticle-num}
\bibliography{ist}

\end{document}